\documentclass[11pt,headings=big,numbers=noenddot,DIV=14,a4paper]{article}%
\pdfoutput=1

\usepackage[english]{babel}
\usepackage[nosort,compress]{cite}
\usepackage{graphicx,epsfig,subfigure}
\usepackage[dvipsnames]{xcolor}
\usepackage[linktoc=page,bookmarks=false,colorlinks=false,
linkbordercolor=RoyalBlue,citebordercolor=ForestGreen,urlbordercolor=CornflowerBlue]{hyperref}
\usepackage[normal,font=small,labelfont=bf,labelsep=period]{caption}
\usepackage{amsmath,amssymb,amsfonts,array,enumerate,slashed,multirow,fancybox,marvosym,wasysym,soul}
\usepackage[normalem]{ulem}

\usepackage[margin=1in]{geometry}

\font\tenrsfs=rsfs10 at 12pt

\font\sevenrsfs=rsfs7
\font\fiversfs=rsfs5
\newfam\rsfsfam
\textfont\rsfsfam=\tenrsfs
\scriptfont\rsfsfam=\sevenrsfs
\scriptscriptfont\rsfsfam=\fiversfs
\def\mathscr#1{{\fam\rsfsfam\relax#1}}

\font\teneusm=eusm10 at 12pt
\newfam\eusmfam
\textfont\eusmfam=\teneusm

\font\tenbbm=bbm10 at 12pt
\newfam\bbmfam
\textfont\bbmfam=\tenbbm

\numberwithin{equation}{section}
\definecolor{darkblue}{rgb}{0,0.2,0.6}
\definecolor{viola}{rgb}{.5,0,.5}
\definecolor{verde}{rgb}{0,.45,0}

\title{\vspace{-2cm}
\vspace{1cm}
\bf\LARGE Higgs and flavour as doors to new physics}
\author{\large Filippo Sala}
\date{\small {\it  Institut de Physique Th\'eorique, Universit\'e Paris Saclay, CNRS, CEA, F-91191 Gif-sur-Yvette, France}}

\begin{document}

\begin{titlepage}
\maketitle

\thispagestyle{empty}

\begin{abstract}
A natural solution to the hierarchy problem of the Fermi scale motivates signals of New Physics at current and near-future experiments. After a critical synthesis of this general motivation, we concentrate our attention on the interplay between LHC searches for new resonances, and precision measurements of both Higgs couplings and flavour violating observables. We do so for i) the Higgs sectors of the NMSSM and MSSM, as paradigmatic examples of theories providing extra scalars, and for ii) CKM-like flavour symmetries, with a focus on $U(2)^3$. This article is mainly based on several papers by the author, but it also reviews other recent related results. Its goal is to provide a synthetic -- yet comprehensive orientation on these subjects, at the dawn of several (ATLAS and CMS, LHCb, NA62 etc.) forthcoming experimental results.%

\end{abstract}


\tableofcontents
\end{titlepage}

\section{Introduction}
\label{intro}
We have \textit{experimental and observational proofs} that the Standard Model (SM) plus General Relativity cannot constitute the final theory of Nature, most notably the existence of Dark Matter, neutrino oscillations and matter/antimatter asymmetry. Unfortunately, none of these proofs gives a precise clue as to where to look for such physics ``beyond the Standard Model''. Perhaps more frustrating than that, there is no guarantee that the solutions of these SM problems are within the reach of the current and near-future experiments. 
From a theoretical point of view, the Standard Model is also self-consistent up to the Planck scale \cite{Buttazzo:2013uya}, and one can engineer solutions to the above experimental problems that do not spoil this consistency nor give observable signals at operating machines (see \textit{e.g.} \cite{Salvio:2015cja}). The mere existence of such solutions poses a problem to the common notion of scientific progress, since it threatens the \textit{falsifiability} of theories in this domain of Science. Other theoretically-driven criteria for ``progress'' of course exist, like reduction of the input parameters, but relying only on them would make the process of scientific inquiry much weaker. 

These solutions of the SM experimental problems, however, do not address at all some of its \textit{theoretical issues}, like the SM flavour problem, the strong CP problem, charge quantisation and the lack of gauge-couplings unification. While not constituting a proof for the need of a new theory, these issues are 
at odds with the taste for simplicity and elegance of most physicists\footnote{\label{foot:Nature_taste}Of course, we do not know if Nature shares the taste of most physicists.
}
.
Among the various theoretical reasons to go beyond the SM, there is one that is supported by more than a taste for beauty, the so-called ``hierarchy problem of the Fermi scale'' (HP) \cite{Wilson:1970ag,'tHooft:1979bh}. In fact, if the Higgs boson $h$ (or any other scalar) couples non-negligibly to some very heavy degree of freedom, then its mass $m_h$ is expected to be of the order of that high scale, unless the parameters at that high energy are ``finely-tuned'' to give the measured value of $m_h \simeq 125$ GeV. In other words, the physics at the Fermi scale would strongly depend on what happens at much shorter distances. This is in contrast with \textit{reductionism}, at least in a strict sense: in the past, much of the scientific progress was possible thanks to the fact that the details of chemistry did not influence biology, those of particle physics did not influence chemistry, etc.. It is in this sense that the HP is more than a desire for elegance and unity. 
However, lack of reductionism does not imply lack of progress in this field, and on the contrary from the measured value of $m_h$ one can constrain theories at much higher energies. Moreover, accidental tunings do exist in Nature, and non-reductionists approaches have proven successful in other domains of Science.

Despite the last warnings, it is logical to use the hierarchy problem to classify model building beyond the SM. This is motivated by the need to (dis)prove such models, and the HP in fact provides the only general indication for New Physics (NP) within experimental reach. It is then useful to restate it in a slightly more precise way, and to synthesise the attitudes proposed to address it.
\subsection{The hierarchy problem and its solutions}
A coupling $y$ of the Higgs $h$ with a fermion $f$, like $y \bar{f} f h$, induces a running of the Higgs mass with the energy $\mu$. This running implies that, to obtain the measured value $m_h(\mu = m_h) \simeq 125$ GeV, the initial value $m_h^2(\mu > m_f)$ has to be chosen with a relative precision --or ``fine-tuning''-- of at least one part in $\Delta_{\rm FT} \sim y^2 m_f^2$. If $m_f \gg m_h$, and $y \sim 1$, then the value of the Fermi scale depends on the details of what happens at much higher energies. This is true, in general, for any degree of freedom coupled to $h$. The attitudes proposed so far to address the HP can be classified in four groups
:
\begin{enumerate}
\item \textbf{Standard Naturalness}.

One can address the HP by adding some new physics at a scale $\Lambda_{\rm NP}$, that screens the sensitivity of $m_h$ to anything above $\Lambda_{\rm NP}$. This can be achieved via a change of regime of the theory or via a symmetry, as done in composite Higgs models and weak-scale Supersymmetry respectively. The fine-tuning left is then of the order of one part in $\sim \Lambda_{\rm NP}^2/m_h^2$. For the Fermi scale to be ``natural'', some NP has then to lie as close as possible to it.
This approach towards the HP has steered theoretical physicists for the last 40 years, not only because it favours NP within experimental reach, but also because it leaves more freedom to solve the other SM problems above $\Lambda_{\rm NP}$.

The lack of experimental evidence for NP, already at LEP, has challenged this ``standard'' approach to the HP \cite{Barbieri:2000gf}, and motivated the pursuit of alternatives. The case for the latters has strengthened after the conclusion of the 8 TeV run of the LHC, that found a SM-like Higgs boson and nothing else.
On top of that, precision measurements have confirmed the SM predictions in a lot of different ways, most notably in flavour physics, electroweak precision tests (EWPT), and measurements of the Higgs couplings. This fact pushes indirectly $\Lambda_{\rm NP}$ to values much higher than the weak scale, unless some model building is worked out.

\item \textbf{Finite Naturalness}.

Sticking to naturalness, having a $\Lambda_{NP}$ close to the Fermi scale is not the only possibility. One can in fact make such NP very weakly coupled to the SM. This leaves a potential issue from gravity, which is expected to give a contribution to $m_h$ of the order of the Planck mass $M_{\rm Pl}$, \textit{i.e.} a fine tuning of one part in $\sim 10^{34}$.
However, a complete quantum theory of gravity is missing, and it is tempting to assume that gravity poses no problems in this respect\cite{Yoon:2002nt,Shaposhnikov:2007nj,Farina:2013mla}\footnote{For attempts to realise such a case see \cite{Dubovsky:2013ira,Salvio:2014soa}.}. Building models under such constraints is very challenging though, and it looks hard, if not impossible, to solve all the SM experimental and theoretical problems within this approach \cite{Giudice:2014tma}. One of the reasons for this difficulty is that, in the absence of a protection mechanism --like Supersymmetry or compositeness--, one has to keep under control any new physics that enters at higher energies.

\item \textbf{Selection on a Multiverse}.

The third way to the HP is to abandon the requirement of a low tuning of the weak scale. After all, another hierarchy problem exists in the current theory of Nature, for which no ``natural'' solution is known: the one of the vacuum energy\footnote{Recently it has been pointed out that it could be solved by protecting the vacuum energy with a ``sequestering mechanism'', see \cite{Kaloper:2015jra} and references therein. Notice that the sequestering mechanism does not solve the hierarchy problem of the Fermi scale.}. In the standard cosmological model, that vacuum energy is responsible for the accelerated expansion of the universe, and its measured value induces a tuning, at the Planck scale, of one part in $\sim 10^{120}$.

Within this perspective, it is still possible to restore some predictive (or postdictive) power for the values of the measured parameters, even without knowing the details of the much higher scale that determines them. This is achieved by assuming that such parameters belong to a large ensemble,  a ``Multiverse'', where they are allowed to span over many possible values (see ref. \cite{Wilczek:2013lra} for a short review with references). A selection mechanism is then thought of, on such an ensemble, to favour some regions with respect to others. The most explored one is the so-called ``anthropic'' mechanism: fundamental parameters have their measured values, because if they differ by little our lives as observers would not be possible at all. In this way, in 1987 Weinberg predicted a value for the vacuum energy \cite{Weinberg:1987dv}, which turned out to be a factor of 10$\div$100 larger than the measured one: an impressive improvement if compared to its 120 orders of magnitude of tuning. A decade later, it was pointed out that also the weak scale can be bounded by similar lines of reasoning \cite{Agrawal:1997gf}. Today this direction is being actively pursued, see \textit{e.g.} the recent \cite{Hall:2014dfa} for an anthropic explanation of the values of the light quarks and electron masses.

\item \textbf{Cosmological relaxation}.

A fourth way to the hierarchy problem explains the small value of the Higgs mass with dynamical evolution in the early universe, and solves the HP without the need of new states close to the electroweak scale \cite{Graham:2015cka}. This can be achieved just adding to the SM a QCD axion and an inflation sector, and we refer the reader to ref. \cite{Graham:2015cka} for explicit models. To avoid possible confusion here we just stress that, unlike the ``finite naturalness'' solutions of group 2, this attitude towards the HP does not need the assumption that gravity poses no problems. This fourth proposal is in its very early days, and we expect a rich activity aimed at its developement, from both the model building and the phenomenological points of view.
\end{enumerate}
%
%
%
%
%
%
%
%
\subsection{Motivation and outline}
The message of high-energy and high-intensity experiments is pretty clear: we have a theory, the Standard Model, that is confirmed to a remarkable level of accuracy in all its predictions, with a very few exceptions. We also have no clear indication of the scale where the needed NP should manifest itself. This poses a challenge to the falsifiability of new theories of Nature and, consequently, to the development of a strategy for the future of the field (which machines should be built, given the limited amount of fundings available?).
Demanding the Fermi scale to be natural still provides the main reason to expect NP to be close in energy. However, different people have different views for the amount of tolerable tuning
, so that the indication for a NP scale at which to aim is not so neat.
It is then of fundamental importance to look for other indications of a NP scale, as well as to probe the Fermi scale and its proximities.

This paper is devoted to the second program, and it concentrates on Higgs and flavour physics. 
It is based mainly on work we performed in those contexts but, at the same time, it aims at giving a wider overview of its subjects. We hope it can provide a useful orientation among the many phenomenological signatures, in Higgs and flavour physics, of a few natural scenarios.
%

The exposition is organised as follows.
Sect. \ref{sec:NMSSM} deals with the next-to-minimal supersymmetric standard model (NMSSM). Sect. \ref{sec:NMSSM_motivation} introduces it as a most natural model, and sect. \ref{sec:NMSSM_parameters} provides a useful parametrisation of its Higgs sector, that makes it easy to discuss the constraints from Higgs coupling measurements. Sects. \ref{sec:NMSSM_Hdec} and \ref{sec:NMSSM_Sdec} discuss the direct and indirect phenomenology of two limiting cases, with sect. \ref{sec:NMSSM_Sdec} covering also the MSSM Higgs sector. Sect.\ref{sec:NMSSM_summary} summarises the messages to be learnt from this study, especially in light of searches at current and future experiments.
Sect. \ref{sec:flavour} deals with flavour symmetries of the quark sector that are close to the Cabibbo Kobayashi Maskawa (CKM) picture of flavour violation, \textit{i.e.} with $U(3)^3$ and $U(2)^3$. After an introduction to the NP flavour problem, we discuss CKM-like symmetries and build the relevant theoretical frameworks in sect. \ref{sec:U2_model}. Their phenomenology, from a general point of view, is then discussed in sect. \ref{sec:U2_pheno}. Sect. \ref{sec:U2_SUSY} deals with the realisation of $U(2)^3$ (and $U(3)^3$) in Supersymmetry, and the consequent interplay of NP searches at the LHC and at flavour factories. 
A partial summary is then given in sect. \ref{sec:U2_summary}.
Finally, general conclusions are exposed in sect. \ref{sec:conclusions}.


\section{The NMSSM Higgs sector}
\label{sec:NMSSM}
Can new Higgs bosons be the first new particles to give signals at experiments?
This general question is very important in relation to the discussion of the previous section. It looks in fact hard to find an anthropic justification for another ``light'' uncoloured scalar, while such particles are present in several models that keep the Fermi scale almost-natural.
The NMSSM is a paradigmatic example of one of these models, and the motivation of this section is to provide a general strategy to look for its CP-even Higgses, at current and future machines.

\subsection{The NMSSM as a most natural model}
\label{sec:NMSSM_motivation}
Weak scale supersymmetry (SUSY, see ref. \cite{Martin:1997ns} for a review) can perhaps be considered as the best candidate to solve the hierarchy between the Fermi and the Planck scales. In its minimal realisation, the minimal supersymmetric Standard Model (MSSM), the Higgs sector constitutes of two doublets, $H_u$ and $H_d$, with vacuum expectation values (vevs) $v_u$ and $v_d$ respectively. The mass of the SM-like Higgs $h$, observed at the LHC8 (here and in the following, we denote by LHC$x$ the LHC run with an energy of $x$ TeV), is bounded by
\begin{equation}
\label{mh_MSSM}
m_h^2 < m_Z^2 \cos^2{2 \beta} + \Delta_t^2,
\end{equation}
where $\tan\beta = v_u/v_d \equiv t_\beta$, and $\Delta_t^2$ is a radiative supersymmetry-breaking contribution, mainly coming from the top-stop sector of the model. Given the measured value of $m_h$, $\Delta_t$ has to be larger than about 85 GeV, \textit{i.e.} of the same size of the SUSY-preserving contribution. This value for $\Delta_t$ implies that the mass of the lightest stop $m_{\tilde t}$ has to be heavier than $\sim 1.5$ TeV, up to $\sim 10 $ TeV if the mixing between the stops is not maximal \cite{Vega:2015fna,Bagnaschi:2014rsa} (previous computations of $m_h$ allowed for stop masses down to $0.8\div 1$ TeV, see e.g. \cite{Hahn:2013ria}). This constitutes an issue from the point of view of tuning, because of the relations
\begin{equation}
\frac{d v^2}{d m_{H_u}^2} \sim \frac{4}{g^2}, \qquad \delta m_{H_u}^2 \sim -\frac{3 y_t^2}{4 \pi^2} m_{\tilde{t}}^2 \log \frac{\Lambda}{m_{\tilde{t}}},
\label{tuning_MSSM}
\end{equation}
where $\Lambda$ is the scale at which SUSY-breaking is mediated. One reads in eq. (\ref{tuning_MSSM}) that the heavier the stop, the worse the tuning of the weak scale $v \simeq 246$ GeV, where the tuning is defined with the logarithmic derivative with respect to the fundamental parameters, \`a la Barbieri-Giudice \cite{Barbieri:1987fn}.  Gluinos also play a major role in the tuning issue, given their large contribution to the stop mass parameter.

The NMSSM alleviates this problem, adding an extra singlet $S$, so that the superpotential $\mathcal{W}$ of the theory reads
\begin{equation}
\mathcal{W} = \mathcal{W}_{\rm MSSM} + \lambda H_u H_d S + f(S),
\label{superpotential}
\end{equation}
 with $f$ a polynomial up to order three in the new field, and $\lambda$ a dimensionless coupling. 
 For a large enough $\lambda$, the importance of this addition is twofold:
 \begin{itemize}
 \item[$\diamond$] the upper bound on the Higgs mass becomes
 \begin{equation}
\label{mh_NMSSM}
m_h^2 < m_Z^2 \cos^2{2 \beta} +\frac{\lambda^2 v^2}{2} \sin^2{2 \beta} + \Delta^2,
\end{equation}
where $\Delta^2$ includes $\Delta_t^2$ and other possible radiative contributions typical of the NMSSM. The dominant contribution to $m_h$ can now be SUSY-preserving, relaxing the Higgs mass constraints on the stop and gluino sectors of the theory;
 \item[$\diamond$] the different dependence of the weak scale $v$ in the Lagrangian parameters allows, for a fixed amount of tuning, to accomodate stop and gluinos parametrically larger by a factor of $O(\lambda/g)$ with respect to the MSSM \cite{Barbieri:2006bg,Hall:2011aa}.
 \end{itemize}
While a very large $\lambda$ is favoured by the second point, it overshoots the measured value of $m_h$, so that a tuning is reintroduced to bring $m_h$ back to 125 GeV \cite{Gherghetta:2012gb} \footnote{\label{foot:tanbeta}This would not be the case for a quite large $\tan\beta$, however $\tan\beta \gtrsim 5$ is disfavoured by EWPT, for $\lambda \gtrsim 1$ and natural values of the Higgsino masses \cite{Gherghetta:2012gb}.}.
The above parametric considerations have been made precise in the thorough study of Ref. \cite{Gherghetta:2012gb}, which found that, for a tuning of 5\% (with a conservative definition of tuning, that multiplies the one in $m_h$ with the one in $v$, thus ignoring their correlation) the lightest new particles are expected to be the scalar Higgses and the lightest neutralino(s) (for other studies of the tuning see e.g. ref.~\cite{Cao:2014kya}). Ref. \cite{Gherghetta:2012gb} also found that the same amount of tuning is consistent with stops at $\sim 1.2$ TeV and gluinos at $\sim 2.5$ TeV. These values are potentially out of reach at the 13 TeV run of the LHC, and make the search for phenomenological signals of the extra Higgs bosons, to which we turn from the next section, more interesting.
We conclude here with the remark that a $\lambda \gtrsim 0.75$ becomes non perturbative before the unification scale $\Lambda_{\rm GUT}\simeq 10^{15\div16}$ GeV \cite{Espinosa:1991gr}, calling for some extra ingredient before $\Lambda_{\rm GUT}$. It is possible however to think of completions that do not spoil unification of gauge couplings, see \textit{e.g.} the one proposed in \cite{Barbieri:2013hxa}, and references therein.

\subsection{A useful parameterization}
\label{sec:NMSSM_parameters}
The aim of this section is to show how to obtain expressions --first derived in ref. \cite{Barbieri:2013hxa}-- for the mixings among the scalars, that are independent of the details of the function $f$ in the superpotential of eq. (\ref{superpotential}). These expressions allow to understand the implications of the Higgs couplings measured at the LHC for the NMSSM, independently of the specific model defined by $f$.

The particle content of the NMSSM Higgs sector, originating from the three fields $H_u$, $H_d$ and $S$, consists in
\begin{itemize}
\item[$\diamond$] three CP-even states $h$, $\phi$ and $H$, of which $h$ is identified with the Higgs boson of mass 125.1 GeV \cite{Aad:2015zhl},
\item[$\diamond$] two CP-odds states $A$ and $A_s$,
\item[$\diamond$] a charged state $H^\pm$.
\end{itemize}
From now on we concentrate on the neutral CP-even sector. For later convenience, we rotate from the gauge eigenstate basis to the one where only one field, $h^0 = \cos{\beta}\,H_u^0 - \sin\beta\,H_d^0$, takes vacuum expectation value (vev) $v$. Let $H^0$ be its orthogonal field, i.e. the neutral component of the doublet that takes no vev, and let $s^0$ be the CP-even component of the scalar $S$. Let $\mathcal{M}^2$ be the $3\times3$ mass matrix in the basis $(H^0,h^0,s^0)$, which is related to the mass basis $(H,h,\phi)$ via 
\begin{equation}
\left(
\begin{array}{c} H  \\ h  \\ \phi \end{array}
\right) = R 
\left(
\begin{array}{c} H^0  \\ h^0  \\ s^0 \end{array}
\right)
, \qquad R = R^{13}_\sigma R^{23}_\gamma R^{12}_\delta,
\label{mixing_angles}
\end{equation}
where $R^{ij}_\theta$ is a rotation of angle $\theta$ in the $ij$ plane. It is then convenient to write the equation for the diagonalization of the mass matrix in the following form

\begin{equation}
\label{mass_diagonalization}
\mathcal{M}^2 = \left(
\begin{array}{ccc}
 M^2(t_\beta,\lambda,\Delta^2,m_A^2) \; &  \begin{matrix}  v M_1  \\ v M_2 \end{matrix}  \\
  \begin{matrix}  v M_1 & &v M_2 \end{matrix} &  M_3^2
\end{array}
\right) = R \left(
\begin{array}{ccc}
 m_H^2 & 0 & 0 \\
 0 & m_h^2 & 0 \\
 0 & 0 & m_\phi^2
\end{array}
\right) R^T,
\end{equation}
where $M_{1,2,3}$ are related to the specific form of $f$, $M^2$ is the upper left $2\times2$ submatrix of $\mathcal{M}^2$ (where we have specified the free parameters on which it depends), and
\begin{equation}
\label{massesAHpm}
m_A^2 = m_{H^\pm}^2 - m_W^2 + \lambda^2v^2/2
\end{equation}
is a diagonal entry of the pseudoscalar mass matrix before diagonalization, with $m_{H^\pm}^2$ the physical mass of the charged Higgs bosons.
Let us now consider the equation for the upper left $2\times2$ matrices in (\ref{mass_diagonalization}). By solving it, one can express the three mixing angles $\delta$, $\gamma$ and $\sigma$, as a function of free parameters that are either physical masses, or in any case independent of $f$ and with a clear physical meaning:
\begin{equation}
\label{mixing_angles}
m_\phi^2, \quad m_H^2, \quad m_{H^\pm}^2, \quad \tan\beta, \quad \lambda, \quad \Delta^2\,.
\end{equation}
The expressions for the mixing angles as a function of the parameters in (\ref{mixing_angles}) are rather lengthy, and we refer the interested reader to refs. \cite{Barbieri:2013hxa,Barbieri:2013nka} for their explicit form.

\begin{figure}[t]
\centering
\resizebox{0.46\textwidth}{!}{%
\includegraphics{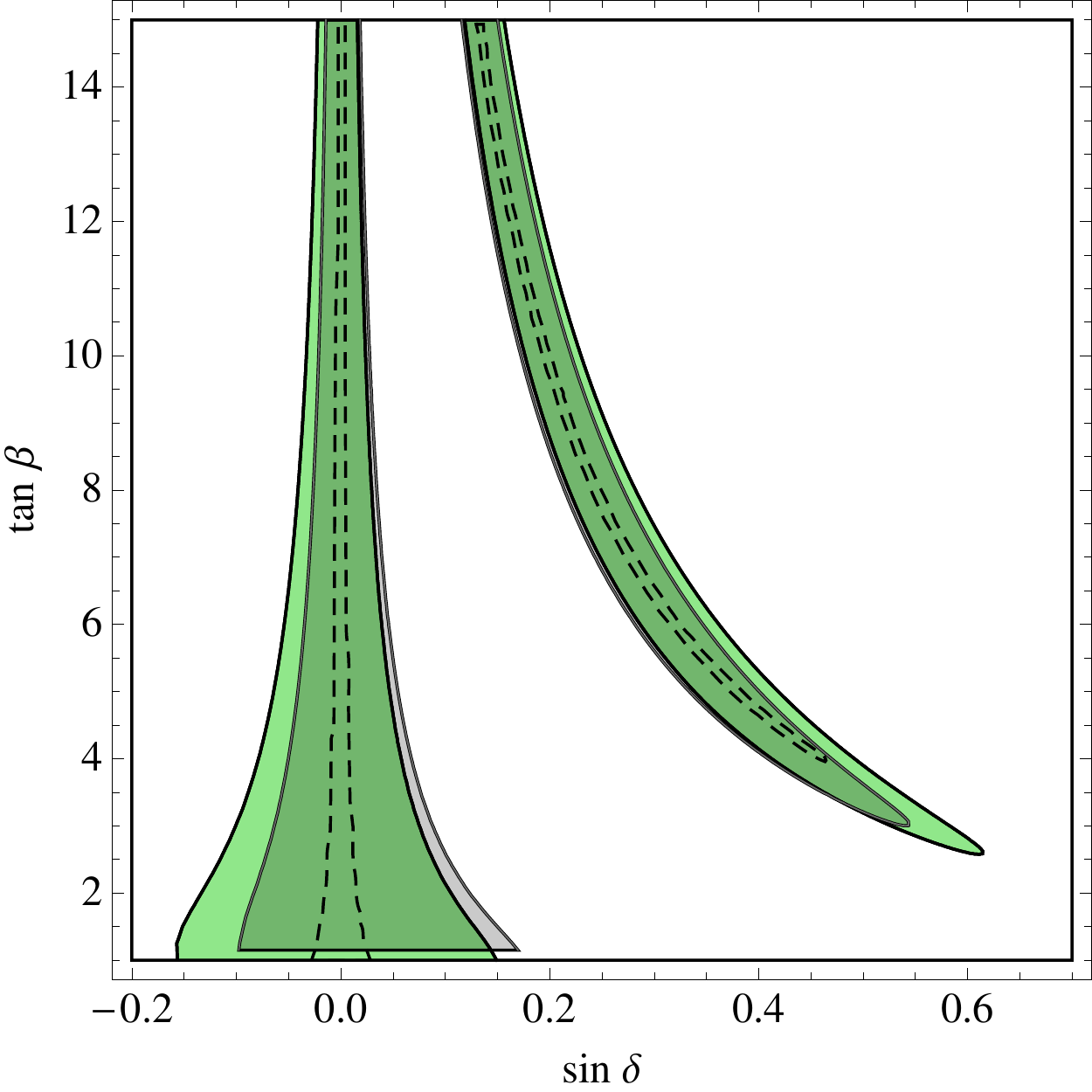}}\hspace{.4 cm}
\resizebox{0.455\textwidth}{!}{%
\includegraphics{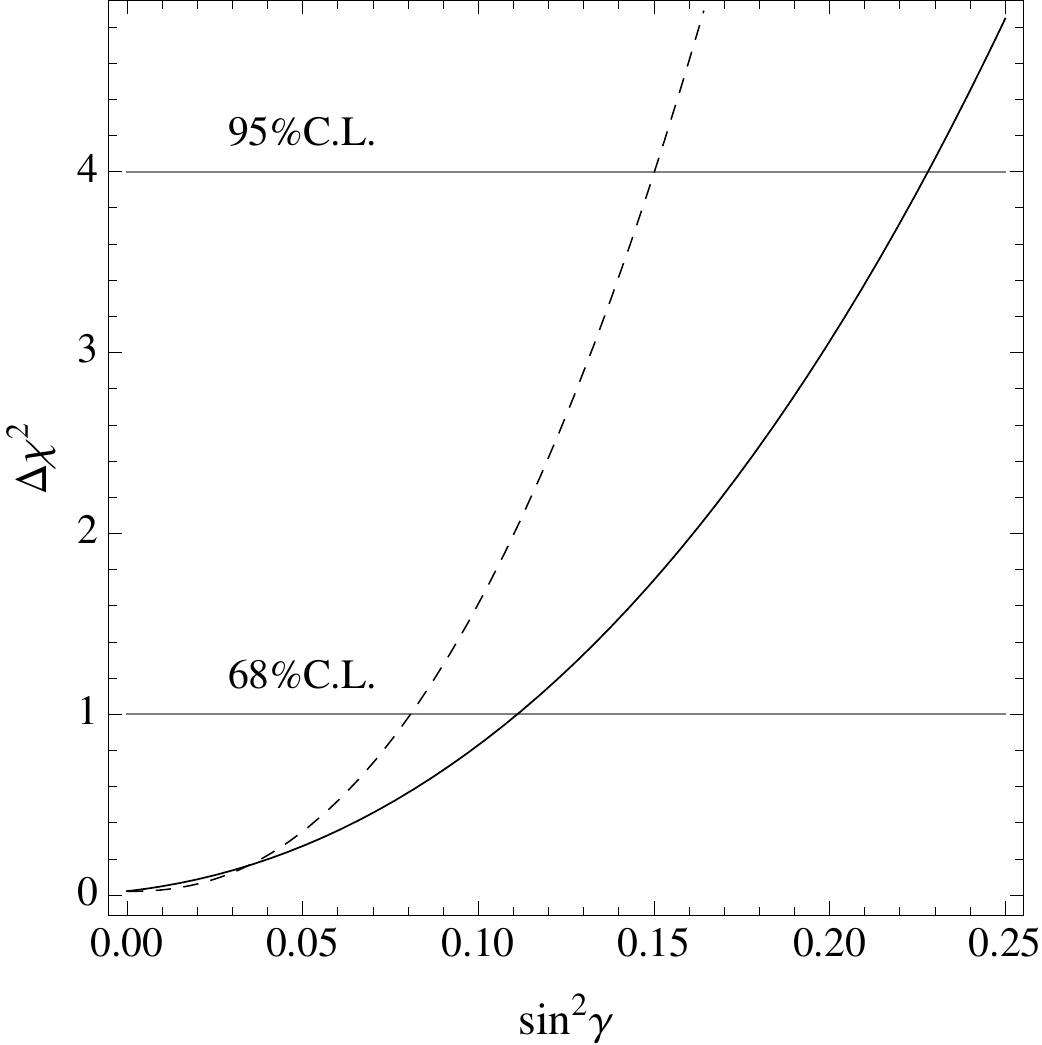}}
\caption{\label{fig:Higgs_fit} Fit of the Higgs signal strengths measured at the LHC8. Left: 3-parameter fit of $s_\delta$, $t_\beta$ and $s^2_\gamma$. The 95\%CL allowed regions are given for $s_\gamma^2 = 0$ (light green) and 0.15 (gray). Dashed lines are contours of the expected allowed region if the LHC14 will not observe deviations in the Higgs signal strengths, with 300 fb$^{-1}$ of luminosity. Right: fit of $s_\gamma^2$ for $\delta = 0$ (continuous), expected LHC14 fit in the same hypothesis of the left-hand plot (dashed). Figure taken from ref. \cite{Andreazza:2015bja}.}
\label{fig:1} 
\end{figure}

\paragraph{Fit of the Higgs couplings.}
The tree-level couplings of the SM-like Higgs boson to up and down quarks of all generations ($u$ and $d$) and to vector bosons $V$ read
\begin{equation}
\label{Higgs_couplings}
\frac{g_{hu\bar{u}}}{g_{hu\bar{u}}^{\rm SM}} = c_\gamma(c_\delta +\frac{s_\delta}{\tan\beta}), \qquad \frac{g_{hd\bar{d}}}{g^{\rm SM}_{hd\bar{d}}}= c_\gamma(c_\delta -s_\delta \tan\beta ), \qquad \frac{g_{hVV}}{g^{\rm SM}_{hVV}}=  c_\gamma c_\delta\,,
\end{equation}
where we have used the shorthand notation $c_\theta,s_\theta = \cos\theta, \sin\theta$. A combined fit to the LHC8 Higgs couplings measurements \cite{Barbieri:2013hxa} yields the results shown as continuous lines in fig. \ref{fig:Higgs_fit}. 
The mixing of the LHC Higgs with the other doublet, $\delta$, is constrained at the 95\% CL by $s_\delta \lesssim 0.15$ for $t_\beta = 1$, down to $s_\delta \lesssim 0.05$ for larger values of $t_\beta$ (the more recent fit of ref. \cite{Craig:2015jba} finds slightly more stringent constraints). On the contrary, a large singlet component is still allowed in the LHC Higgs, the 95\% CL constraint on the mixing being $s_\gamma < 0.48$. Concerning upcoming experiments, if the LHC14 will not observe deviations in the Higgs couplings, then a quite large singlet component would still be allowed, while the mixing $\delta$ would be constrained down to a few percent level. This can be seen from the dashed lines of fig. \ref{fig:Higgs_fit}, see ref. \cite{Barbieri:2013nka} for details.\footnote{The ``banana region'' on the left-hand plot of fig. \ref{fig:Higgs_fit}, allowing larger values of $s_\delta$, comes from the fact that both signs are still allowed for the $h$ coupling to bottom quarks.} The constraints and projections for each angle are derived setting the other one to zero, but they do not change much if this assumption is relaxed, as one can see in the left-hand side of fig. \ref{fig:Higgs_fit}.

Let us now define two limiting cases, where different degrees of freedom are relevant:
\begin{itemize}
\item[a)] $h$ plus $\phi$. The doublet-like state $H$ is decoupled, $m_H \gg m_\phi, m_h$, and $\delta,\sigma \ll \gamma$. This implies also $m_{H^\pm} \gg m_\phi, m_h$, and the free independent parameters are $m_\phi, \lambda, t_\beta, \Delta$.
\item[b)] $h$ plus $H$. The singlet-like state $\phi$ is decoupled, $m_\phi \gg m_H, m_h$, and $\gamma,\sigma \ll \delta$. The free independent parameters are now $m_H, t_\beta, \Delta$; $\lambda$ and $m_{H^\pm}$ are functions of them.
\end{itemize}
To discuss the phenomenology, since our aim is to give a general search strategy, we stick to the case a) and b) as outlined in sect. \ref{sec:NMSSM_Hdec} and sect. \ref{sec:NMSSM_Sdec} respectively.

\subsection{An extra singlet-like Higgs}
\label{sec:NMSSM_Hdec}
In the limiting case a), the mixing angle between the SM-like state and the singlet reads
\begin{equation}
\label{singamma}
s_\gamma^2 = \frac{c^2_{2 \beta} m_Z^2 + s^2_{2 \beta} \lambda^2 v^2/2 + \Delta^2 -m_h^2}{m_\phi^2 - m_h^2},
\end{equation}
and it governs alone most of the phenomenology. The couplings to SM particles of the two states are in fact given by
\begin{equation}\label{singlet_couplings}
\frac{g_{hff}}{g_{hff}^{\rm SM}} =  \frac{g_{hVV}}{g_{hVV}^{\rm SM}} =  c_\gamma, \qquad  \frac{g_{\phi ff}}{g_{hff}^{\rm SM}} =  \frac{g_{\phi VV}}{g_{hVV}^{\rm SM}}   = -s_\gamma\,
\end{equation}
where $VV = WW, ZZ$ and $ff$ stands for any couple of SM fermions.
The signal strengths $\mu_{A\to B} = \sigma_{pp\to A}\times {\rm BR}_{A \to B}$ then are\footnote{We assume that no other channels are open, like $\phi \to A_s A_s$ or $\phi \to $LSP\,LSP.}
\begin{align}
\Delta\mu/\mu_{\rm SM} \equiv \frac{\mu_{h\to {\rm SM}} - \mu_{\rm SM}(m_h)}{\mu_{\rm SM}(m_h)} &= s^2_\gamma,\label{signal_strength_Higgs}\\ 
\frac{\mu_{\phi \to VV,ff}}{\mu_{\rm SM}(m_\phi)} &= s^2_\gamma \times (1-{\rm BR}_{\phi \to hh})\,,\label{signal_strength}\\
\frac{\mu_{\phi \to hh}}{\sigma_{\rm SM}(m_\phi)}&= s^2_\gamma \times {\rm BR}_{\phi \to hh}\,, \label{signal_strength_hh}
\end{align}
where $\sigma_{\rm SM}(m)$ is the total cross section of a SM Higgs boson of mass $m$, $\mu_{\rm SM}(m)$ is its signal strength into the channel of interest (\textit{i.e.} into any channel in the first line, into $VV$ or $ff$ in the second line, and into $hh$ in the third line), and BR$_{\phi \to hh}$ is the branching ratio of $\phi$ into $hh$, which depends on the details of the model.
To display the current and future reaches of direct and indirect searches, we stick to the analysis we performed in ref. \cite{Buttazzo:2015bka}. We give here only the general message, for details (like future sensitivities, how the projections for direct bounds have been derived etc.) we refer the reader to ref. \cite{Buttazzo:2015bka}.
We leave the physical mass $m_\phi$ and $t_\beta$ free to vary, while we fix $\lambda$ and $\Delta$. The interval we choose for $t_\beta$ is motivated by EWPT (see footnote \ref{foot:tanbeta}), and we consider $m_\phi > m_h$ for definiteness. Concerning $\lambda$ and $\Delta$, we consider two possible cases
\begin{itemize}
\item $\lambda = 1.2$ and $\Delta = 70$ GeV. The value of $\lambda$ lies in an interval of minimal tuning of the model, the one of $\Delta$ can be achieved for a mass of the lightest stop from a few hundreds GeV (maximal mixing) to several TeVs. We checked however that the phenomenology depends on $\Delta$ in a very mild way.
\item $\lambda = 0.7$ and $\Delta = 80$ GeV. The value of $\lambda$ is at the edge of the region of perturbativity up to the GUT scale. The value of $\Delta$ is fixed at 80 GeV, since a lower one would make it difficult to reproduce $m_h = 125.1$ GeV in the whole parameter space we consider.
\end{itemize}

\begin{figure}[t]
\centering
\resizebox{0.47\textwidth}{!}{%
\includegraphics{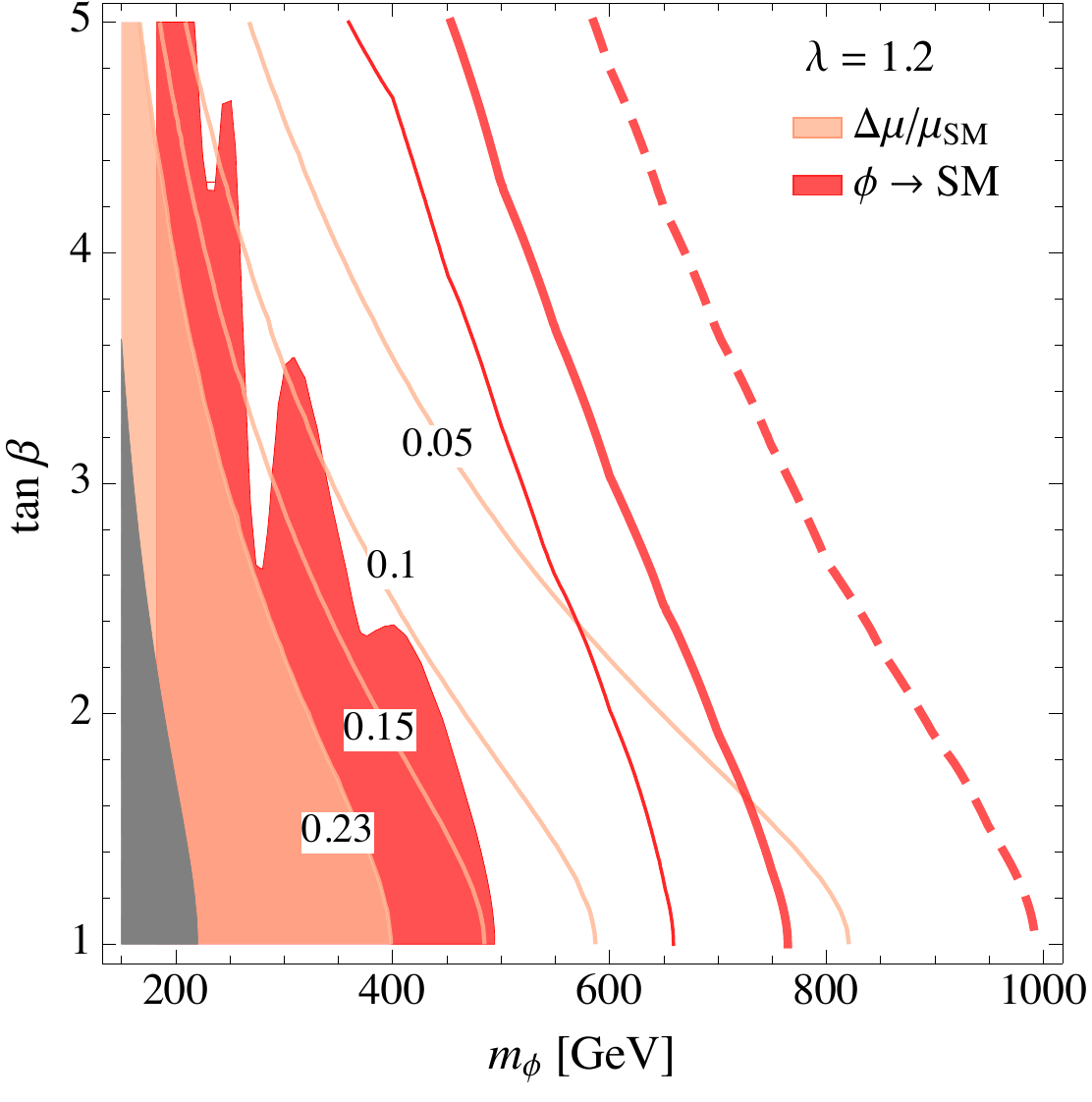}
}\hspace{.4 cm}
\resizebox{0.47\textwidth}{!}{%
\includegraphics{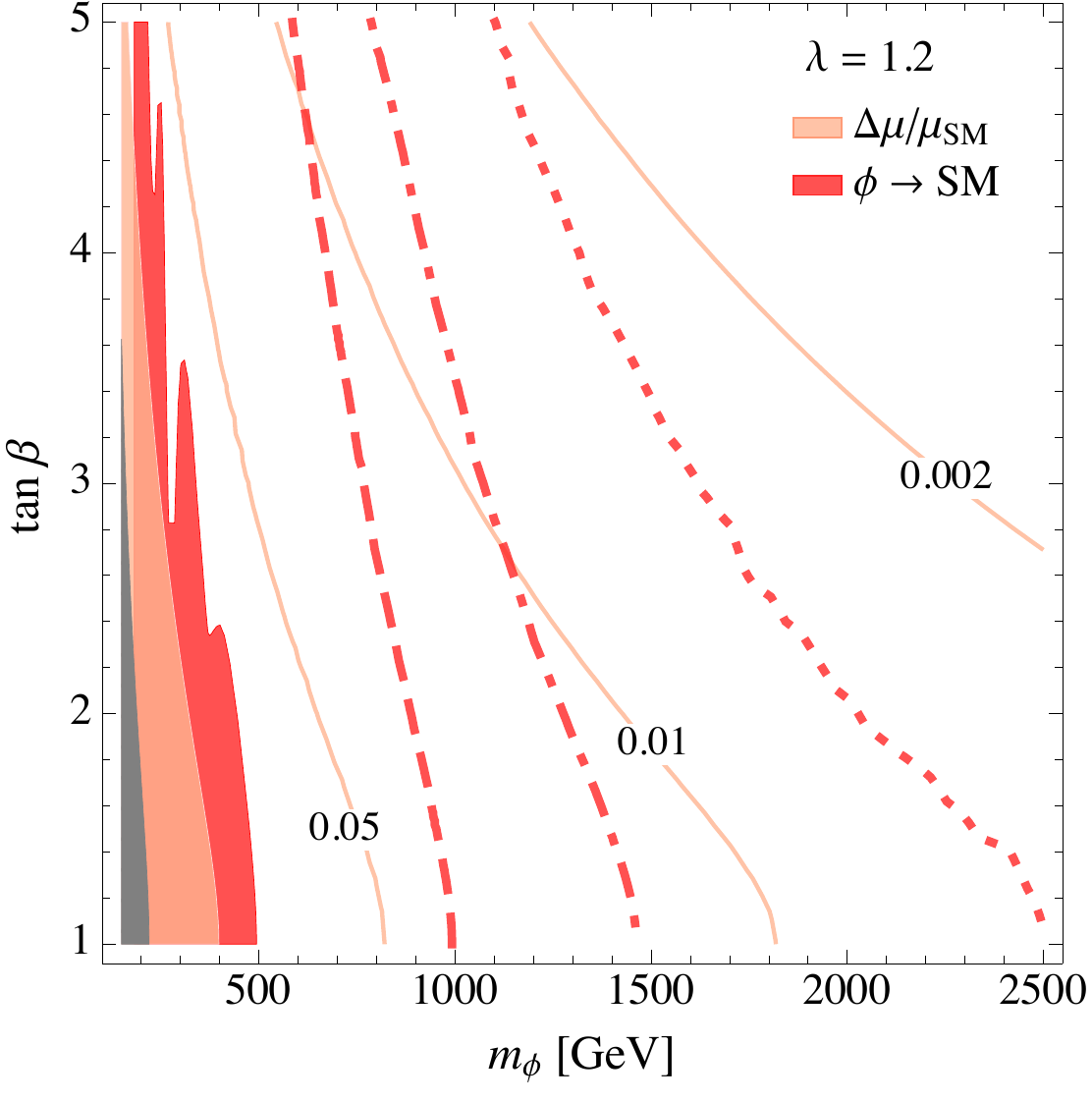}}
\caption{Case of $\lambda = 1.2$ and $\Delta = 70$ GeV. Shaded areas: 95\% CL exclusions from Higgs coupling measurements (pink) and searches for $\phi \to {\rm SM}$ (red) at the LHC8. Pink lines: contours of $s_\gamma^2$. Red lines: expected reach of the LHC13 (continuous thin), LHC14 (continuous), HL-LHC (dashed), HE-LHC (dot-dashed), and FCC-hh (dotted). BR$_{\phi \to hh}$ fixed to its asymptotic value 1/4. Grey: unphysical regions. Taken from ref.~\cite{Buttazzo:2015bka}.}
\label{fig:NMSSM_l12}       
\end{figure}

\begin{figure}
\centering
\resizebox{0.47\textwidth}{!}{%
\includegraphics{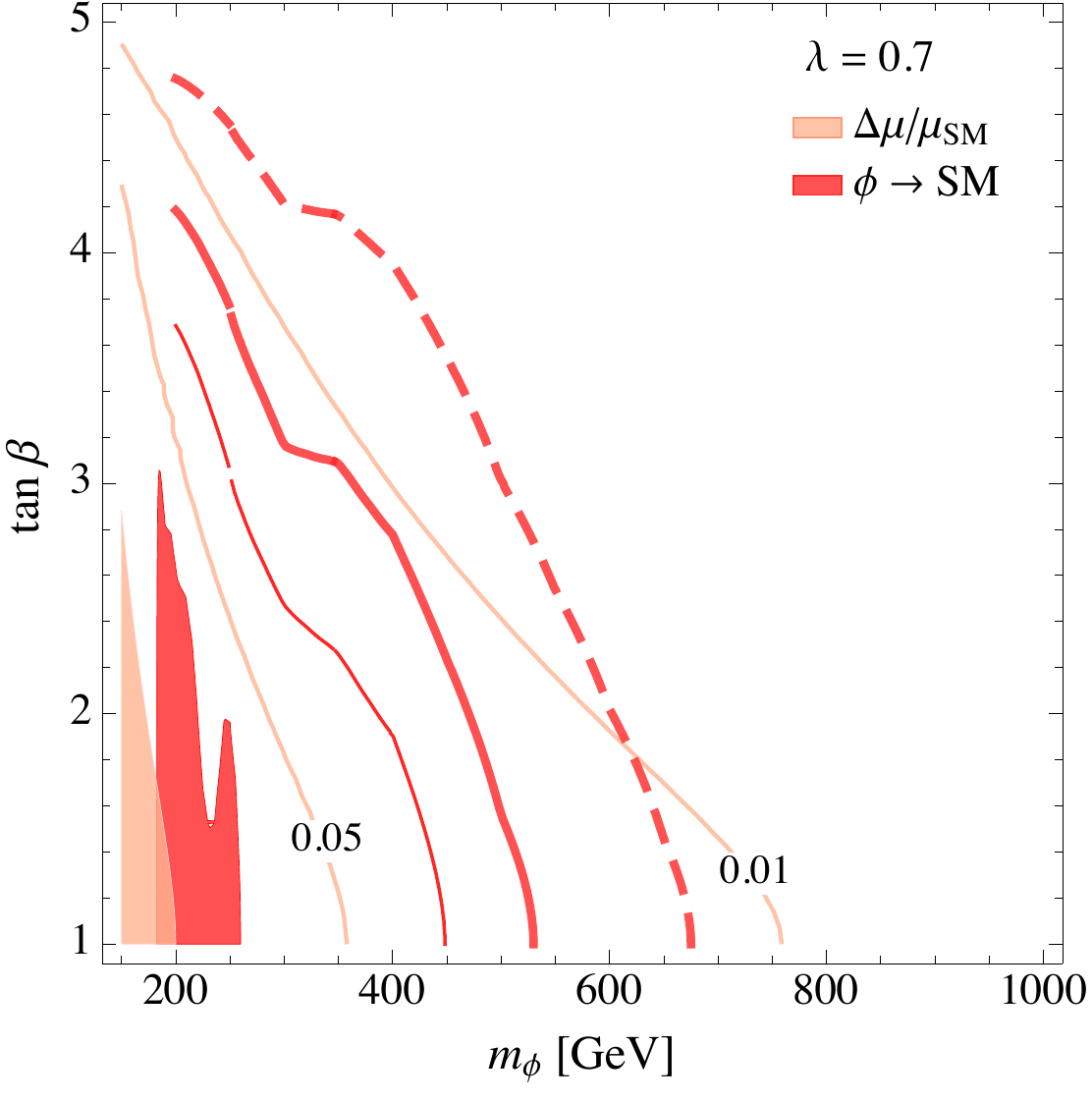}
}\hspace{.4 cm}
\resizebox{0.47\textwidth}{!}{%
\includegraphics{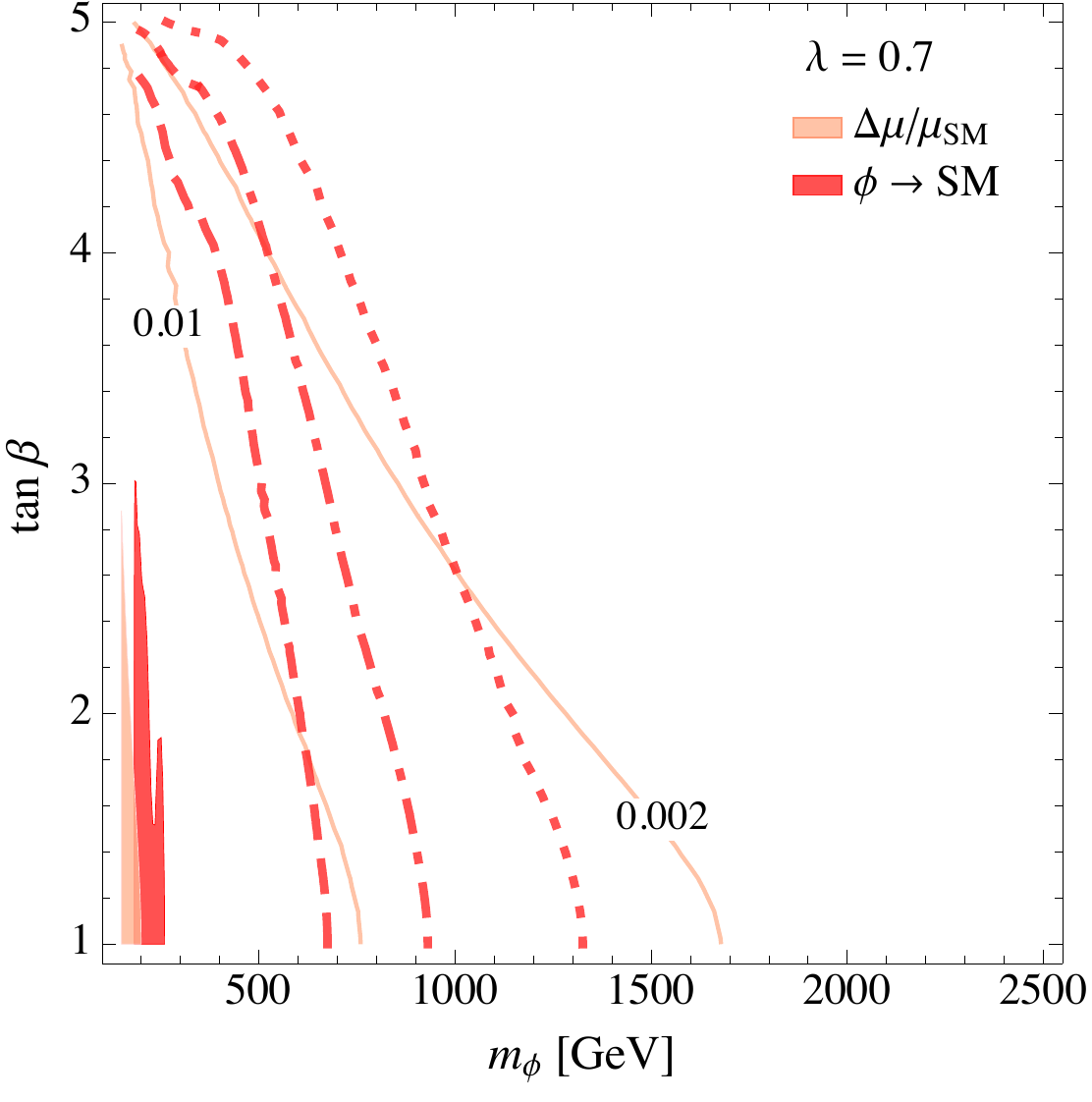}
}
\caption{Case of $\lambda = 0.7$ and $\Delta = 80$ GeV. All the rest as in fig. \ref{fig:NMSSM_l12}. Taken from ref. \cite{Buttazzo:2015bka}.}
\label{fig:NMSSM_l07}       
\end{figure}

\paragraph{Higgs signal strengths vs direct searches.} Eq. (\ref{signal_strength_Higgs}) is sufficient to study the current and future capability of Higgs coupling measurements to probe this limiting case of the NMSSM. In figs. \ref{fig:NMSSM_l12} and \ref{fig:NMSSM_l07}, we show in pink several contour lines of $s^2_\gamma$, of interest for the expected sensitivities of the future LHC stages (left-hand plots), and of future leptonic machines (right-hand plots).
LHC14 with 300 fb$^{-1}$ is expected to be sensitive to values of $s_\gamma^2$ between 0.1 and 0.15 (see also the right-hand plot of fig. \ref{fig:Higgs_fit}), while the high-luminosity run has the potential to go down in $s_\gamma^2$ by an extra 0.05. Future leptonic machines are expected to probe $s_\gamma^2$ at the per-cent (ILC) down to the per-mille (CLIC, CEPC and FCC-ee) level. For more details about future projections, we refer the reader to the Snowmass study \cite{Dawson:2013bba}. The lines of $s^2_\gamma$ provide also a reference to infer the impact of an invisible branching ratio for $h$, the current 95\% CL bound in fact scales as $s_\gamma^2 < 0.23 - 0.78\,\text{BR}_{h \to {\rm inv.}}$ \cite{Barbieri:2013nka}.

It is interesting to compare the reach of Higgs coupling measurements with the one of direct searches for the extra scalar $\phi$.
The largest direct signals are expected to originate from decays of $\phi$ into $ZZ$, $WW$ and $hh$. The more promising fermionic channel is $t\bar{t}$, but it is subdominant with respect to the one in vector bosons, especially at high masses.
To know the signal strengths of $\phi$, the value of BR$_{\phi \to hh}$ is needed. In general, it depends on the details of the potential and on the specific NMSSM model under consideration. It is possible to simplify the discussion by noting that, for $m_\phi \gg m_W$, the following relation holds
\begin{equation}
{\rm BR}_{\phi \to hh} = {\rm BR}_{\phi \to ZZ} = {\rm BR}_{\phi \to WW}/2\,.
\label{BRhh_asymptotic}
\end{equation}
This motivates to fix ${\rm BR}_{\phi \to hh} = 1/4$ when displaying the reach of future searches: besides being a reference value, for large enough $m_\phi$ it gives also an accurate result.
Let us come now to the actual experimental searches. CMS has recently provided a combination \cite{Khachatryan:2015cwa} of various searches for an extra scalar in $ZZ$ and $WW$, with the full luminosity of the concluded LHC runs. This constitutes the stronger available constraint from $VV$ channels. Searches of scalar resonances into the $hh$ channel have also been performed, both in $4b$ \cite{ATLAS4b,Khachatryan:2015yea} and in the $2b2\gamma$ \cite{Aad:2014yja,CMS:2014ipa} final states. The impact of all these searches is shown as shaded areas in figs. \ref{fig:NMSSM_l12} and \ref{fig:NMSSM_l07}.
Among direct searches, for ${\rm BR}_{\phi \to hh} = 1/4$ the reach is largely dominated by the CMS $VV$ combination. This provides stronger constraints than Higgs coupling measurements in the whole region of interest.\\
Concerning now future experiments, we consider as benchmarks the next stages of the LHC (13 TeV with 100 fb$^{-1}$, 14 TeV with 300 and 3000 fb$^{-1}$), as well as a 33 TeV (HE-LHC) and a 100 TeV (FCC-hh) $pp$ colliders, both with 3 ab$^{-1}$ of integrated luminosity.
To obtain an estimate of the reach of such machines, we use a method introduced in \cite{SalamWeiler} and developed in \cite{Thamm:2015zwa}. This method stems from the observation that, for center-of-mass partonic energies $\sqrt{\hat{s}} \gg m_W$, the exclusions in cross-section times branching ratio are driven by the parton luminosities responsible for the backgrounds. The procedure relies on a number of assumptions which, though reasonable, make the reaches that we show only an indicative estimate of what a future machine could do, especially for the case of very-high energy colliders. For a thorough discussion of the assumptions behind this method, and for some checks of its validity, we refer the reader to ref.~\cite{Buttazzo:2015bka}.

Future direct reaches are displayed as red lines in figs. \ref{fig:NMSSM_l12} and \ref{fig:NMSSM_l07}, together with the contour lines of $s_\gamma^2$, that we already discussed. One can appreciate in this way the expected interplay of direct and indirect searches, at various future machines. The most effective way to probe an extra-singlet like state, in the future LHC stages and for $\lambda$ between the two values considered, is clearly via $ZZ$ and $WW$ resonant production. For $\lambda = 0.7$, direct searches at the HL-LHC are expected to do better even than coupling measurements at the per-cent level, in the ballpark of the expected ILC sensitivity. A 100 TeV $pp$ collider (FCC-hh) could be more effective than a per-mille precision in the Higgs couplings, below $m_\phi \sim 1$ TeV. For the case $\lambda = 1.2$ instead, a per-cent level precision in the Higgs couplings would be more constraining than the expected direct reach at the HL-LHC. Analogously, a per-mille level precision in $\Delta\mu/\mu_{\rm SM}$ would probe more parameter space than the FCC-hh. In general, probing the case of a smaller $\lambda$ looks much more challenging.

\paragraph{Trilinear Higgs couplings.}
The phenomenology discussed so far is valid for any scalar potential. However, it does not exhaust the possible signals measurable at experiments, like deviations in the trilinear Higgs coupling. To know their size in the NMSSM with $H$ decoupled, in principle one needs to know the full details of the potential. This would make the predictions of the model less sharp, depending on many parameters. In ref. \cite{Buttazzo:2015bka} it has been shown that this is not the case, and that the parameters that mostly influence the trilinear Higgs couplings $g_{hhh}$ and $g_{\phi hh}$, besides $m_\phi$ and $t_\beta$, are $\lambda$ and the singlet vev $v_s$ (unless $v_s \gg v$). If a $VV$ resonance and/or a $\Delta \mu/\mu \neq 0$ will be observed in the future, by measuring self Higgs couplings one could then obtain specific information about the singlet vev.

We find that, to obtain a large BR$_{\phi \to hh}$, one typically needs a $v_s$ in the range of  tens of GeV. Down to $|v_s| \sim 20$ GeV the theory is still weakly coupled, \textit{i.e.} the width of $\phi$ and the self Higgs coupling are still under control. However, those values of $v_s$ typically require large hierarchies in the Lagrangian parameters and, more importantly from the point of view of phenomenology, they imply very light Higgsinos if one does not allow for a $\mu H_u H_d$ term in the superpotential.
\begin{figure}[t]
\centering
\resizebox{0.47\textwidth}{!}{%
\includegraphics{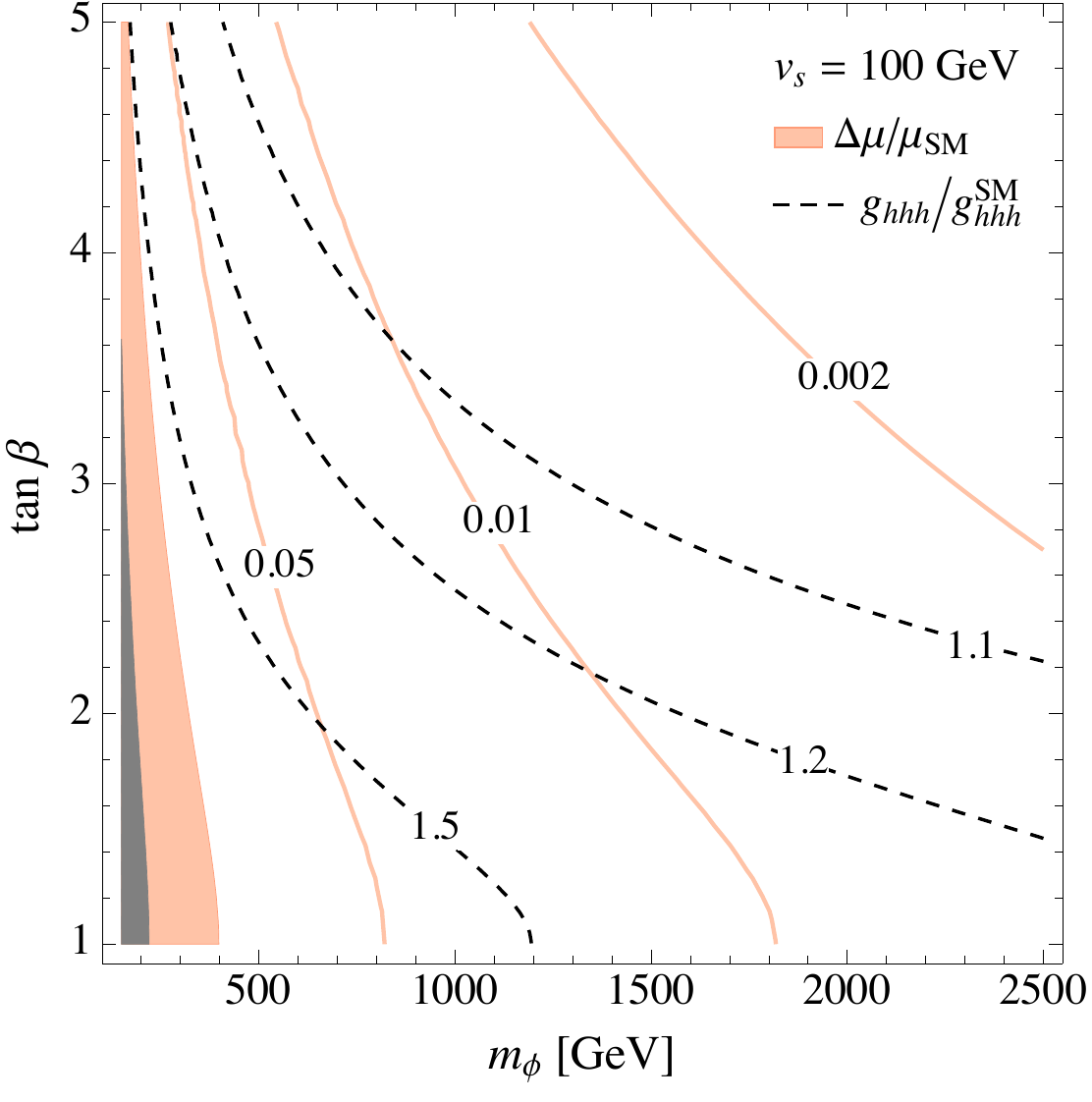}
}\hspace{.4 cm}
\resizebox{0.47\textwidth}{!}{%
\includegraphics{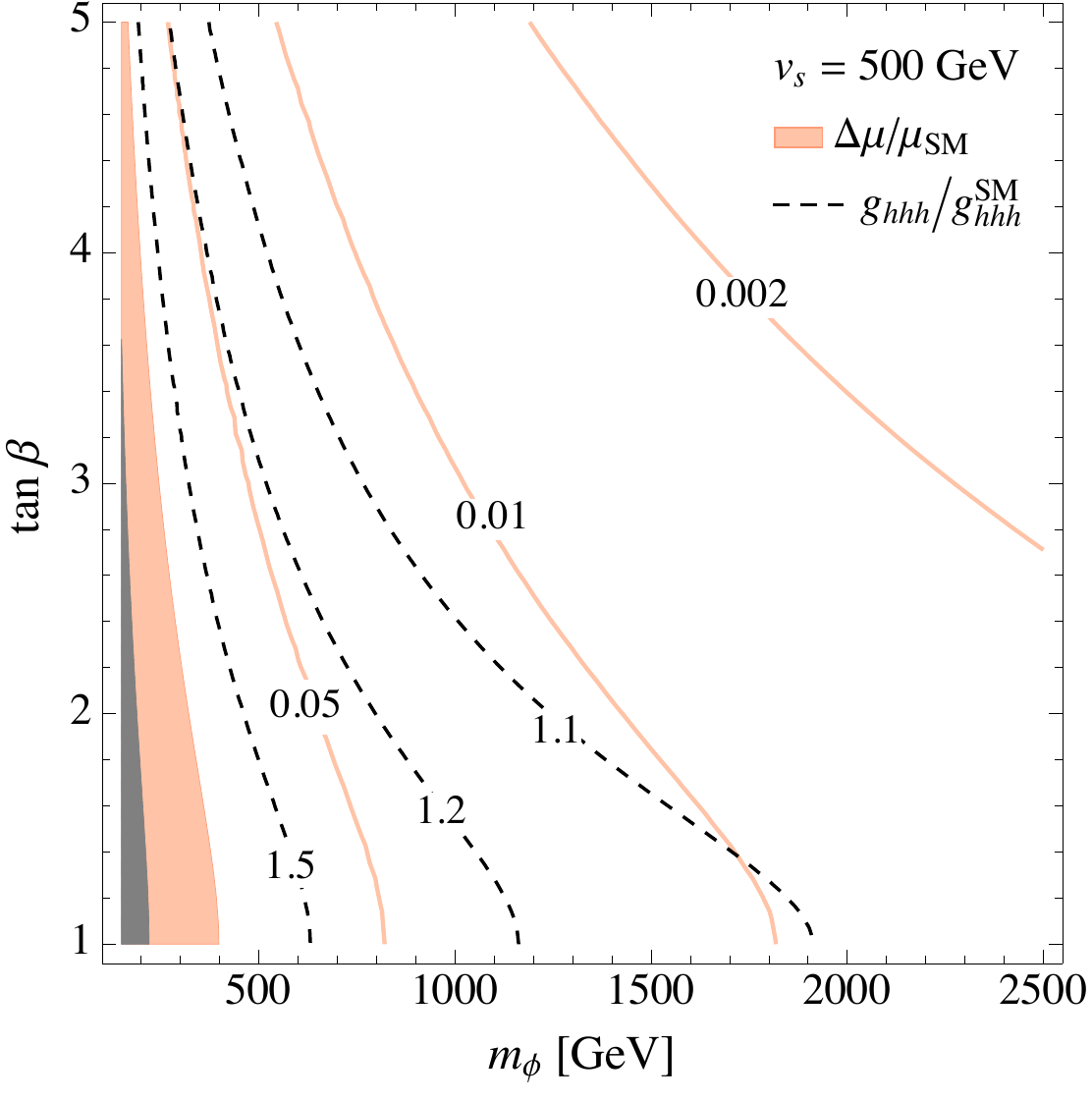}
}
\caption{$\lambda = 1.2$, $\Delta = 70$ GeV, $\kappa = 0.3$, and $v_s = 100$ GeV (left-hand plot), $v_s = 500$ GeV (right-hand plot). Shaded pink: 95\% CL exclusions from LHC8 Higgs couplings measurements. Lines: $s_\gamma^2$ (pink) and $g_{hhh}/g_{hhh}^{\rm SM}$ (black dashed). Grey region: unphysical.}
\label{fig:NMSSM_triple}       
\end{figure}
Thus, in fig. \ref{fig:NMSSM_triple} we show the predictions for $g_{hhh}/g_{hhh}^{\rm SM}$, for $\lambda = 1.2$, and $v_s = 100$ and 500 GeV. The values of the other parameters have little impact, for definiteness we fixed the potential to that of a scale-invariant NMSSM ($f(s) = \kappa S^3$, see ref. \cite{Ellwanger:2009dp} for a review), and the only other free parameter $\kappa$ to 0.3. For reference, we show also contour lines of $s_\gamma^2$. The current and projected reach in the $\phi \to hh (4b)$ are never competitive in the regions of the parameters space we consider here, so that direct searches are dominated everywhere by $VV$, and the previous conclusions --about the interplay of direct and indirect searches-- still hold.
A message of fig. \ref{fig:NMSSM_triple} is that it is possible to observe deviations in the trilinear Higgs coupling before observing them in the Higgs signal strengths, for example at the HL-LHC or at the ILC (for which the expected sensitivity on $g_{hhh}/g_{hhh}^{\rm SM}$ are in the ballpark of 50\% and 20\% respectively \cite{Dawson:2013bba}). While this property is expected from a bottom-up point of view \cite{Pomarol:2013zra}, in the majority of concrete models such large deviations are not allowed \cite{Gupta:2013zza}.
Notice finally that we give results for $\lambda = 1.2$, since for $\lambda = 0.7$ and the same range of $v_s$, deviations in the trilinear Higgs coupling are typically below the 10\% level.

\paragraph{Fully mixed case, and a $\gamma\gamma$ signal from a lighter scalar.}

So far, we have discussed the case of an extra singlet heavier than 125 GeV. The possibility of a lighter singlet-like scalar is open, and its best direct probe still comes from LEP searches of $h \to b\bar{b}$ \cite{Schael:2006cr}, see \textit{e.g.} ref. \cite{Falkowski:2015iwa} for an updated comparison of constraints. Concerning the LHC, the ATLAS collaboration has performed a search for scalar resonances in the $\gamma\gamma$ channel, down to a mass of 65 GeV \cite{Aad:2014ioa}. Its sensitivity to a lighter singlet-like Higgs is still weaker than Higgs coupling measurements, so that it makes sense to ask: if we observe a signal, could we explain it in the NMSSM, within a fully mixed situation?

\begin{figure}
\centering
\resizebox{0.47\textwidth}{!}{%
\includegraphics{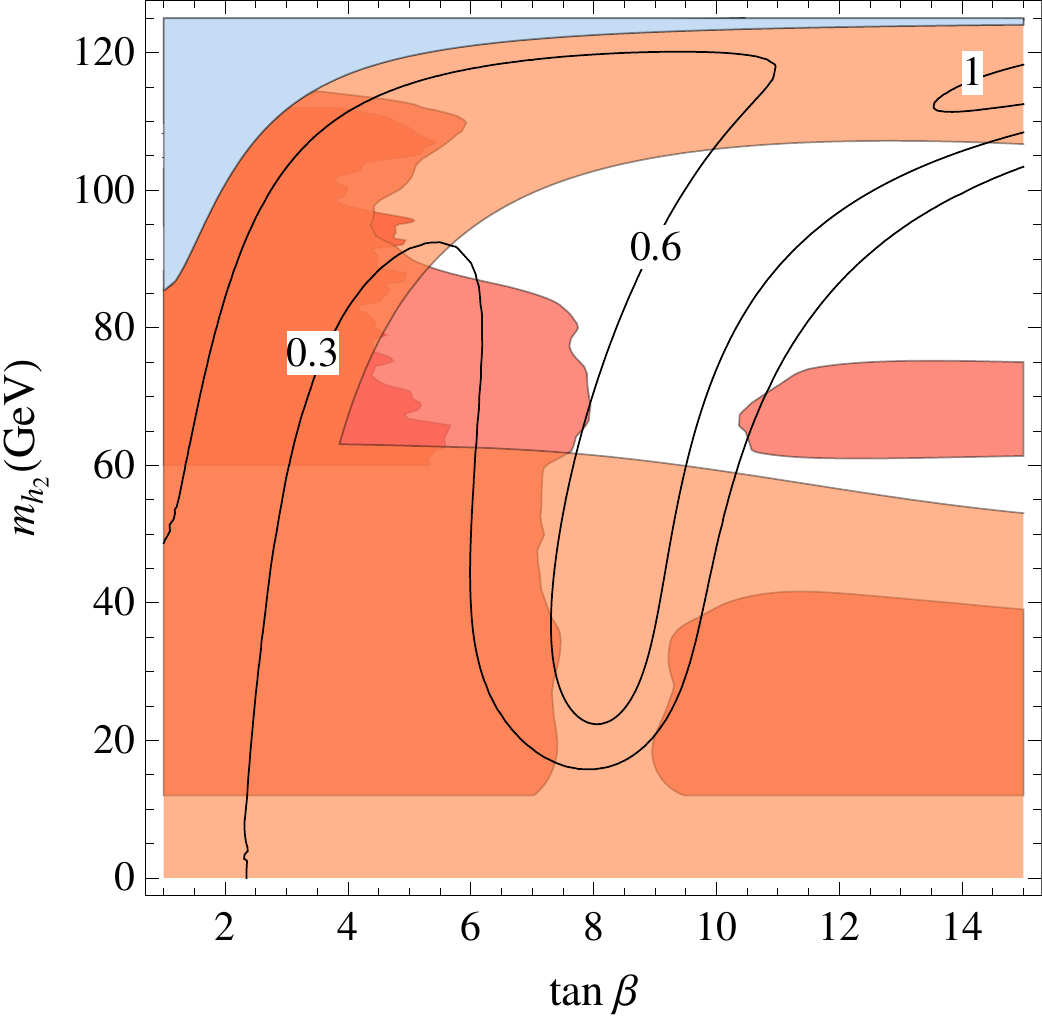}
}\hspace{.4 cm}
\resizebox{0.47\textwidth}{!}{%
\includegraphics{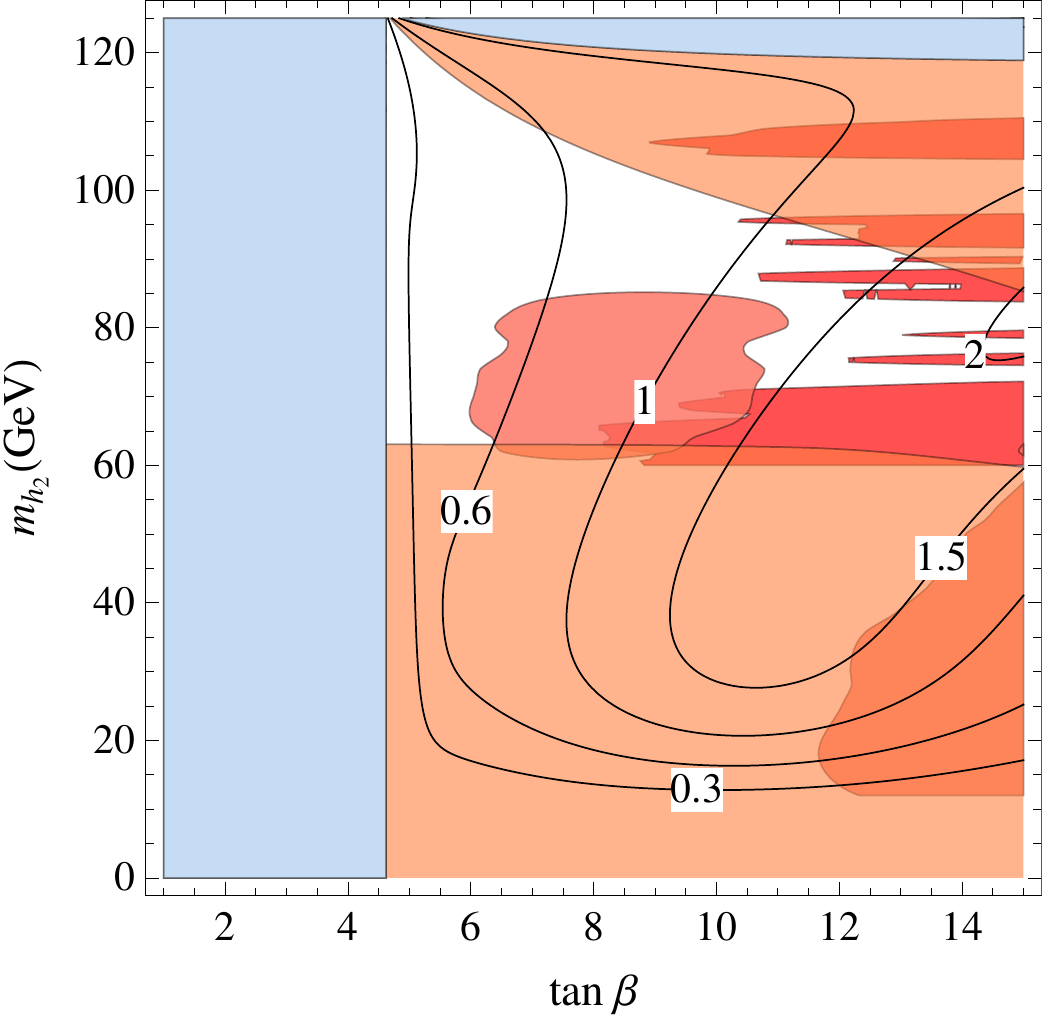}
}
\caption{Shaded areas: 95\% CL exclusions from Higgs coupling measurements at the LHC8 (pink) and searches for $\phi \to b\bar{b}$ (red) and $\phi \to$ hadrons (dark red) at LEP \cite{Schael:2006cr}. Lines: signal strength $\phi \to \gamma\gamma$ normalised to that of a SM Higgs of the same mass. Light blue: unphysical regions. Taken from ref. \cite{Barbieri:2013nka}, where the notation $\phi \to h_2$ was used.}
\label{fig:gammagamma}       
\end{figure}

We give here an example of the use of the fully mixed formalism, where all the angles $\delta, \gamma, \sigma$ of (\ref{mixing_angles}) are different from zero. We show in fig. \ref{fig:gammagamma} contour lines of $\mu_{\phi\to\gamma\gamma}/\mu_{\rm SM}(m_\phi)$, for $\lambda = 0.1$ (see also ref. \cite{Badziak:2013bda} for a study of this case) and $\lambda = 0.8$. We do not study the case of larger $\lambda$ because it does not allow to reproduce $m_h = 125$ GeV for low values of $\tan\beta$. The other parameters are specified to minimise the exclusion coming from LEP searches for $h \to b\bar{b}$ and $h \to$ hadrons, which are shown in red and dark red respectively. The pink shaded regions are instead excluded from Higgs coupling measurements.
Above roughly 80 GeV, the ATLAS search of ref. \cite{Aad:2014ioa} is sensitive to values of $0.5\div1$ for $\mu_{\phi\to\gamma\gamma}/\mu_{\rm SM}(m_\phi)$, and thus it already probes some regions allowed by Higgs coupling measurements. This shows that direct and indirect searches can already be complementary in that mass region, and motivates experimental collaborations to keep looking for scalar resonances lighter than 125 GeV.

\subsection{An extra doublet-like Higgs and the MSSM}
\label{sec:NMSSM_Sdec}

In the limiting case b) of sect. \ref{sec:NMSSM_parameters}, \textit{i.e.} of the singlet-like state decoupled and $\gamma, \sigma \ll \delta$, one has an extra relation between the free parameters of the previous case. All the phenomenology can then be described in terms of just three free parameters, which we choose as $m_H$, $\tan\beta$ and $\Delta$.
Moreover, the ``light'' CP-even degrees of freedom consist now in three scalars, of which two are neutral, $h$ and $H$, and one is charged, $H^\pm$. The mixing angle between the doublets, $\delta$, can be expressed as in eq. (\ref{singamma}), upon the substitution $m_\phi \to m_H$. 
The couplings of the two scalars to SM particles read
\begin{align}
 &\frac{g_{hu\bar{u}}}{g_{hu\bar{u}}^{\rm SM}} = c_\delta +\frac{s_\delta}{\tan\beta}, &
 &\qquad \frac{g_{hd\bar{d}}}{g^{\rm SM}_{hd\bar{d}}} = \frac{g_{h\ell\bar{\ell}}}{g^{\rm SM}_{h\ell\bar{\ell}}} = c_\delta -s_\delta \tan\beta, &
 &\qquad \frac{g_{hVV}}{g^{\rm SM}_{hVV}}=   c_\delta,& \label{hcouplings_Sdec}\\
 &\frac{g_{Hu\bar{u}}}{g_{hu\bar{u}}^{\rm SM}} = s_\delta -\frac{c_\delta}{\tan\beta}, &
 &\qquad \frac{g_{Hd\bar{d}}}{g^{\rm SM}_{hd\bar{d}}} = \frac{g_{H\ell\bar{\ell}}}{g^{\rm SM}_{h\ell\bar{\ell}}} = s_\delta +c_\delta \tan\beta, &
 &\qquad \frac{g_{HVV}}{g^{\rm SM}_{hVV}}=   s_\delta,& \label{Hcouplings_Sdec}
\end{align}
so that the signal strengths of both $h$ and $H$ into fermions strongly depend on $\tan\beta$, not only on the mixing angle.

\begin{figure}[t]
\centering
\resizebox{0.47\textwidth}{!}{%
\includegraphics{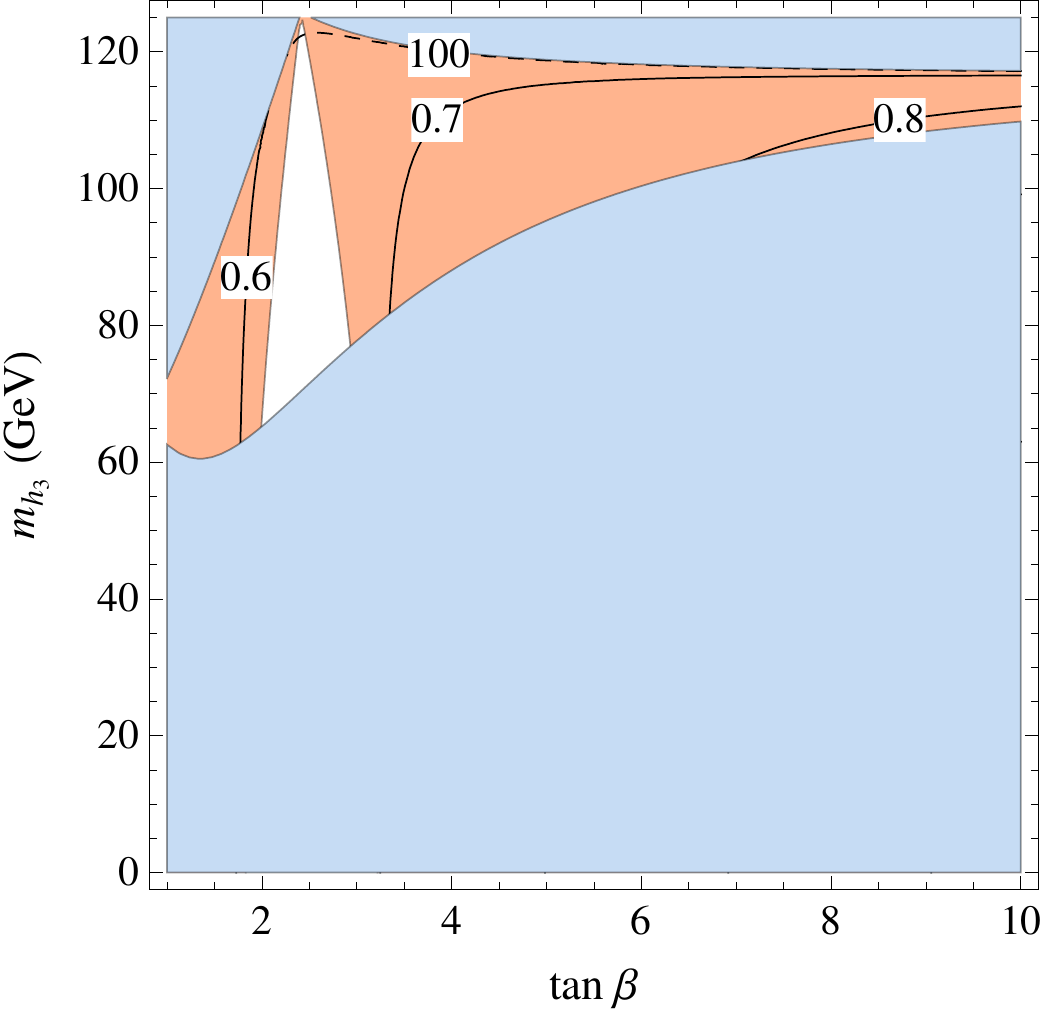}
}\hspace{.4 cm}
\resizebox{0.47\textwidth}{!}{%
\includegraphics{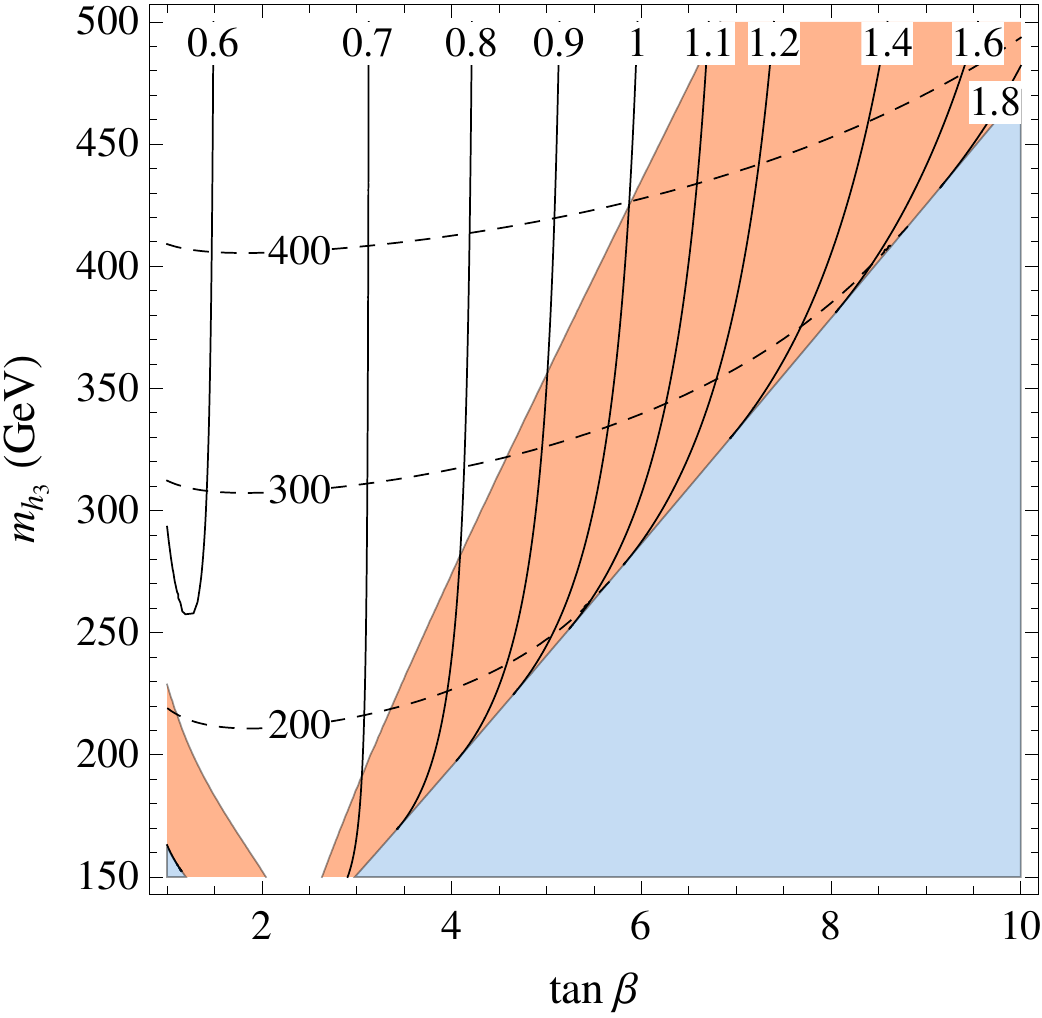}}
\caption{Shaded pink: 95\% CL exclusions from Higgs coupling measurements at the LHC8, for $m_H < m_h$ (left) and $m_H > m_h$ (right). Lines: values of $\lambda$ (continuous), and of $m_{H^\pm}$ (dashed). Light blue: unphysical regions. Taken from ref.~\cite{Barbieri:2013nka}, where the notation $H \to h_3$ was used.}
\label{fig:Sdec_LHC8}       
\end{figure}

\begin{figure}[h!]
\centering
\resizebox{0.47\textwidth}{!}{%
\includegraphics{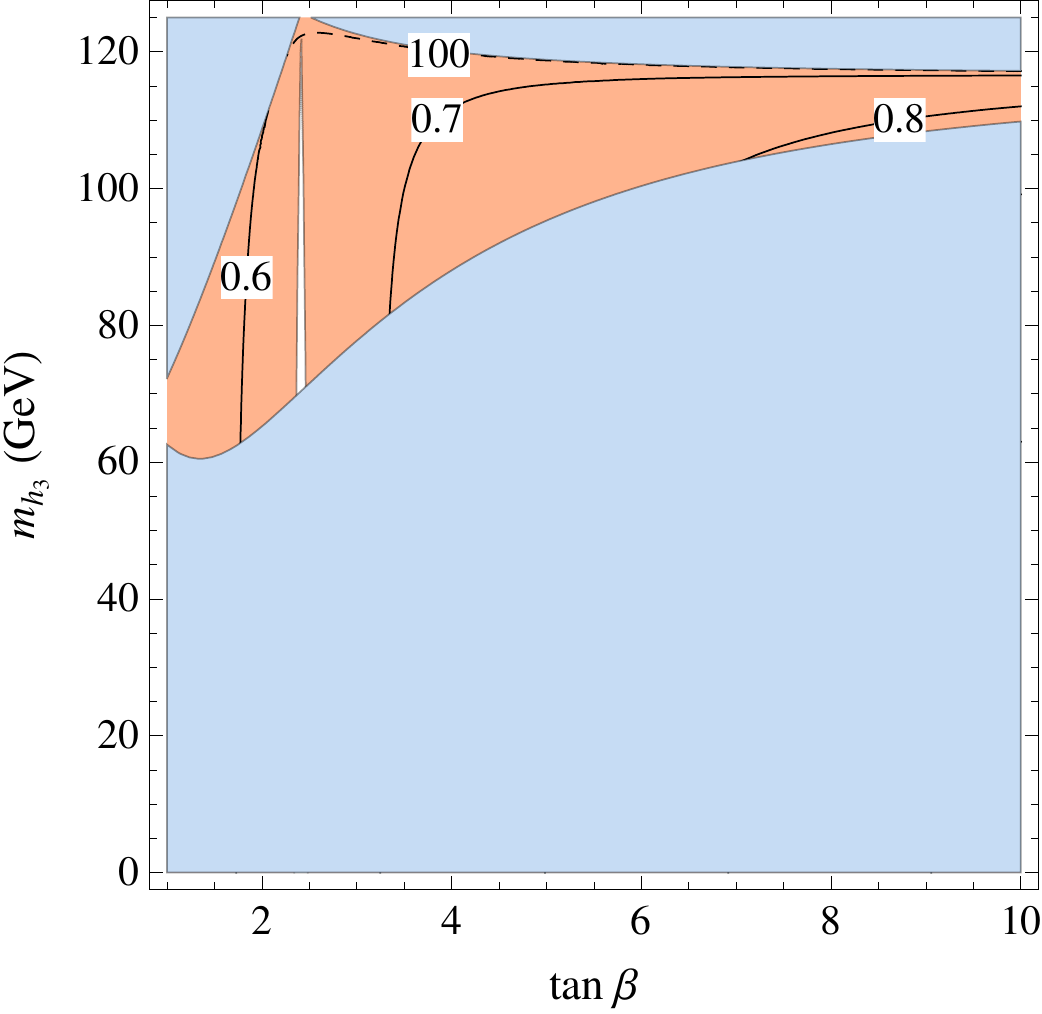}
}\hspace{.4 cm}
\resizebox{0.47\textwidth}{!}{%
\includegraphics{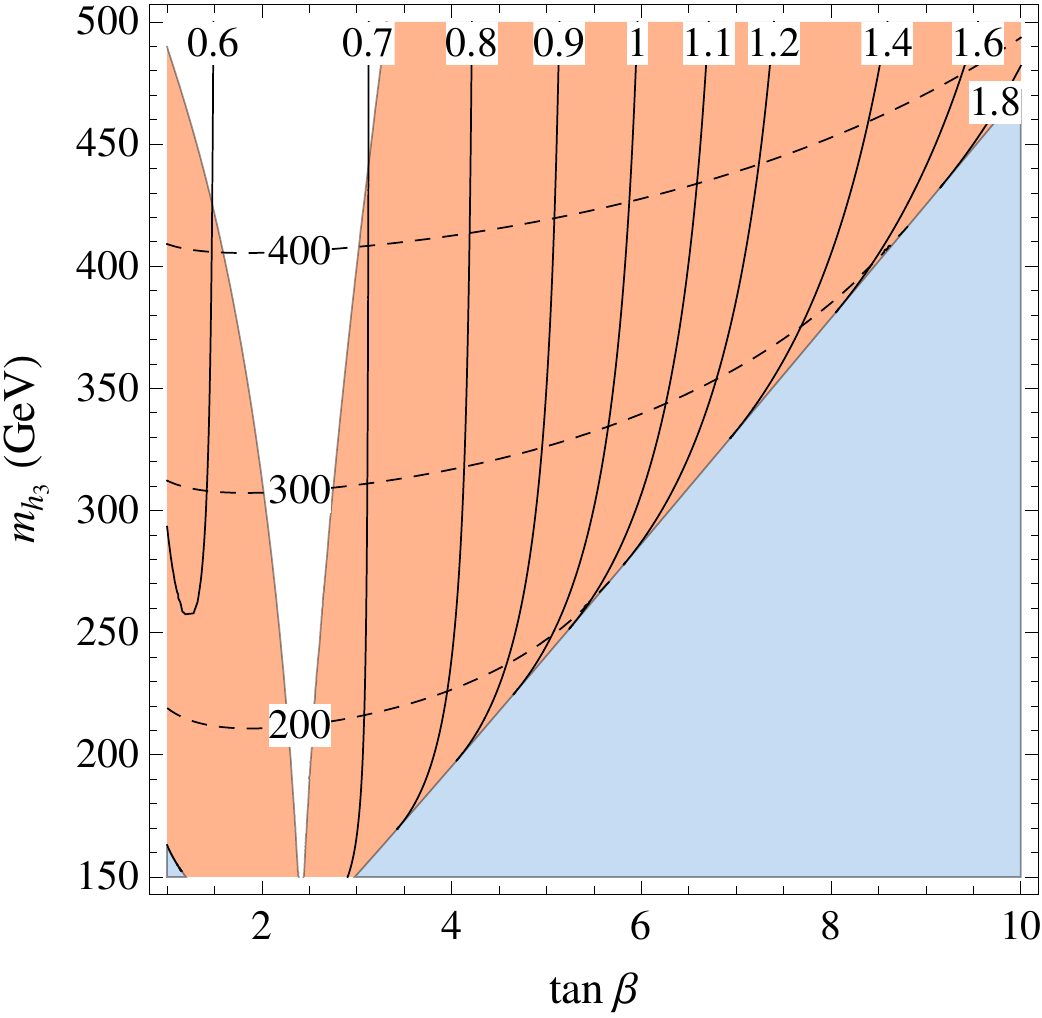}
}
\caption{Shaded pink: expected reach at the LHC14 with 300 fb$^{-1}$, from Higgs coupling measurements. All the rest as in fig. \ref{fig:Sdec_LHC8}. Taken from ref. \cite{Barbieri:2013nka}, where the notation $H \to h_3$ was used.}
\label{fig:Sdec_LHC14}       
\end{figure}

Using eq. (\ref{hcouplings_Sdec}) we draw, in figures \ref{fig:Sdec_LHC8} and \ref{fig:Sdec_LHC14}, the constraints coming from Higgs coupling measurements at the LHC8, as well as the expected reach of the LHC14 with 300 fb$^{-1}$ of luminosity. We do so for both cases of $m_H < m_h$ and $m_H > m_h$. In the first one, a small region of parameter space remains open, and it will be almost completely probed by Higgs coupling measurements \cite{Barbieri:2013nka}. Furthermore, a light $H$ implies a light $H^\pm$ (see dashed isolines in the plot), and flavour measurements like BR$_{B\to X_s \gamma}$ impose $m_{H^\pm} \gtrsim 480$ GeV (95\%CL) \cite{Misiak:2015xwa}, unless a cancellation with contributions from other light sparticles is invoked.
In the case of  $m_H > m_h$, a wider region of parameter space is open, especially at high masses. Still, Higgs coupling measurements at future colliders are expected to explore most of that region. It is interesting that they also ``select'' a value of $\lambda$ between 0.6 and 0.7, irrespectively of the mass ordering. This is due to our choice to fix the radiative contribution $\Delta$ to 75 GeV in both figures, so that a precise (and sizeable) contribution form the $\lambda$ term is needed to reproduce the Higgs mass, in line with our motivation to consider the NMSSM. A different value of the radiative contribution does not strongly affect the phenomenology, unless one goes to $\Delta \gtrsim 85$ GeV. Notice that, for such high values of $\Delta$, a smaller $\lambda$ would be preferred. The phenomenology would then closely resemble that of the MSSM, to which we shall return at the end of this section.
An invisible branching ratio of $h$, \textit{e.g.} into a pair of Dark Matter particles or into two pseudoscalars $A$, would not significantly affect the impact of the fit on $\delta$ and $\tan\beta$, provided BR$_{h\to{\rm inv.}} \lesssim 0.2$ \cite{Barbieri:2013nka}.
Contrary to the $H$ decoupled case of sect. \ref{sec:NMSSM_Hdec}, the trilinear Higgs coupling is consistent with its SM value at a level below $10\div20$\% \cite{Gupta:2013zza}, so that one does not expect to see deviations at the HL-LHC, nor at the lower energy runs of ILC and CLIC (see ref. \cite{Dawson:2013bba} for future sensitivities).

Direct searches of scalar resonances provide a necessary complementary tool to explore the model. As already said, measurements of Higgs signals strengths are expected to be the strongest probe of the parameter space. However, and contrary to the case of a light singlet-like state, the ``alignement'' limit $\delta \to 0$ does not imply that also the couplings of $H$ (and $A$ and $H^\pm$) to SM states go to zero. For an exactly SM-like Higgs $h$, the extra scalar $H$ can still be probed by searches of resonances in up ($t\bar{t}$) and down ($b\bar{b}$) quarks, in leptons ($\ell \bar{\ell}$) and in $\gamma\gamma$\footnote{It is also interesting that, in this limit, the couplings to SM fermions of $H$ and the $A$ coincide.}. This is important since the NMSSM, when $\phi$ is decoupled, realises ``alignment without decoupling'', \textit{i.e.} one can have $\delta = 0$ even for low values of $m_H$ (the line $\delta = 0$ can be inferred from the allowed regions of fig. \ref{fig:Sdec_LHC14}). On the contrary, $\delta = 0$ implies BR$_{H \to VV}$ = BR$_{H \to hh}$ = 0 (= BR$_{ A \to Zh}$ = BR$_{H^\pm \to W^\pm h}$).
For an updated thorough discussion of the direct signals of 
$H$, $A$ and $H^\pm$, and of their interplay with Higgs coupling measurements, we refer the reader to ref. \cite{Craig:2015jba} and to references therein\footnote{The study of ``Type 2 2HDM'' performed in \cite{Craig:2015jba} covers the cases of the MSSM, and of the NMSSM with singlet decoupled.}. Here we just summarise its content, when relevant to our discussion.
\begin{itemize}
\item[$\diamond$] For $\delta = 0$, searches of a scalar or pseudoscalar resonance decaying into $\tau^+ \tau^-$ are currently the only ones able to probe a part of our parameter space, excluding values of $m_{H,A} \lesssim 350$ GeV for $\tan\beta$ very close to one (they barely reach $\tan\beta = 2$). For higher masses, the $t\bar{t}$ channel opens and searches into $\tau^+ \tau^-$ are not effective, leaving a wide region unexplored. For values of $\tan\beta \gtrsim 10$, both gluon fusion and $b\bar{b}$ associated production grow significantly, and searches in the $\tau^+\tau^-$ channel exclude values of $m_{H,A}$ smaller than $450\div550$ GeV.

\item[$\diamond$] Going away from the alignment limit, searches of resonances in the $VV$, $hh$ and $Zh$ channels start to become effective. Their impact, together with the one of $\tau^+ \tau^-$ searches, is summarised in fig. \ref{fig:Craig_current} for a reference value of 300 GeV for $m_H$ and $m_A$ respectively.
\begin{figure}[t]
\centering
\resizebox{0.96\textwidth}{!}{%
\includegraphics{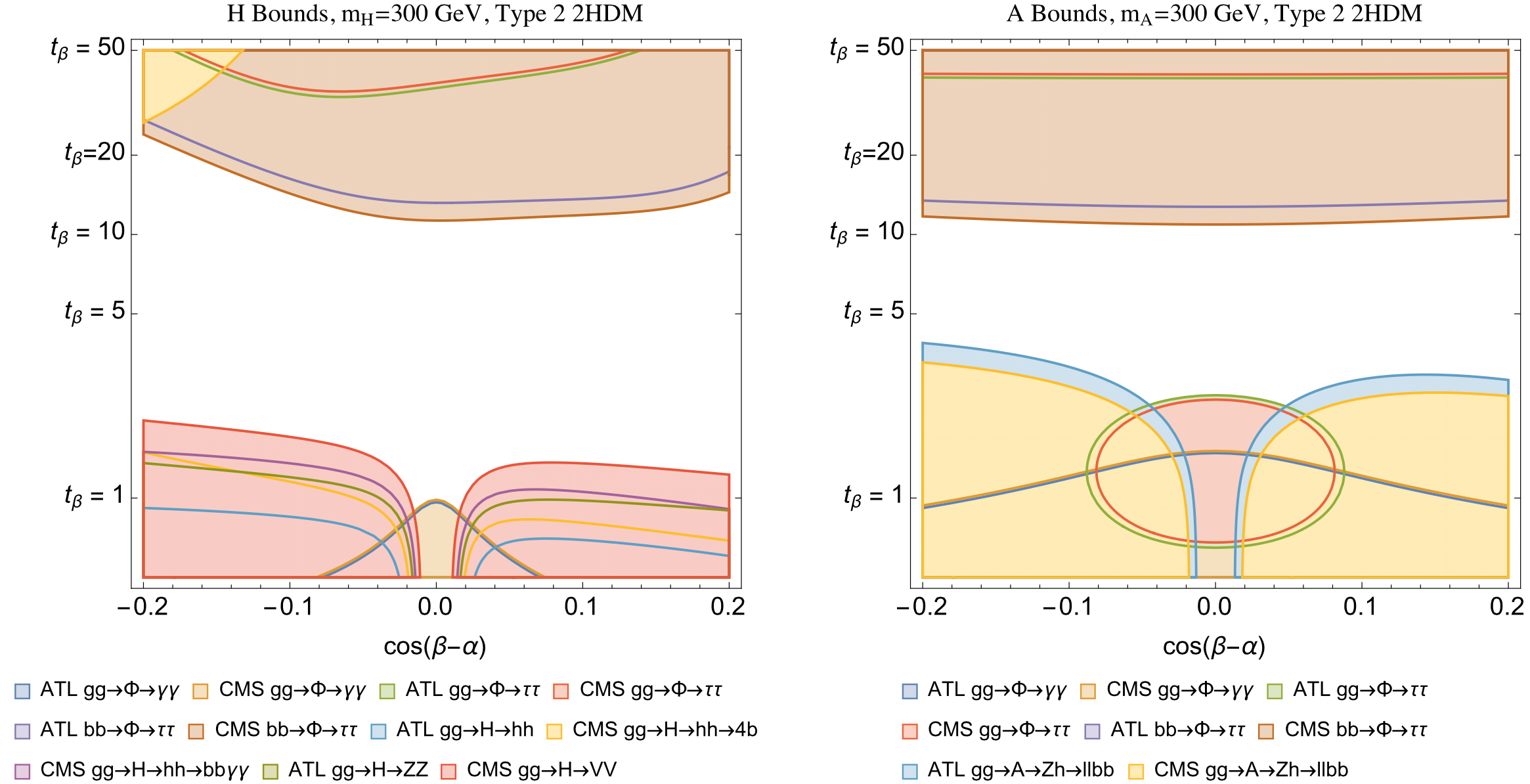}
}
\caption{\label{fig:Craig_current} Shaded: regions excluded by direct searches of $\Phi = H$ (left) and $\Phi = A$ (right) at the LHC8, see legend for color code. Figure taken from ref. \cite{Craig:2015jba}, where $\cos{(\beta-\alpha)} = \sin\delta$.} 
\end{figure}
%
To understand the interplay with Higgs coupling measurements, fig. \ref{fig:Craig_current} should be compared with the left-hand side of fig. \ref{fig:Higgs_fit}. In doing so, one has to keep in mind that $\sin\delta = \cos(\beta - \alpha)$ and that $m_A$ and $m_{H^\pm}$ are related by eq. (\ref{massesAHpm}). Direct searches are more effective than indirect ones only for very small values of $\tan\beta$. Moreover, for $m_{H,A} = 500$ GeV or higher, they are already incapable to probe the region left open by Higgs coupling measurements.

\item[$\diamond$] The above considerations underline the importance to perform searches in the $t\bar{t}$ channel, that is expected to be one of the most important avenues where to look for the extra Higgses of this scenario, provided one finds a way around the expected LHC14 systematics. This channel has the potential to cover a range of small to moderate $\tan\beta$, and values of $\delta$ down to the alignment limit. Another complementary channel, that is expected to cover an analogous region of parameter space, is the associated production of $H^\pm$ with $t\bar{b}$, with subsequent decay $H^\pm \to t\bar{b}$.
Away from the alignment limit, and for small values of $\tan\beta$, searches of $H \to VV$ and $A \to Zh$ will likely play an important role (see \textit{e.g.} \cite{Craig:2013hca} for a study of future sensitivities in the $ZZ$ channel).
\end{itemize}
To our knowledge, the study of the sensitivity of future machines --up to the FCC-hh-- in all the channels mentioned above is still to be completed
.

\paragraph{The MSSM for comparison.}
\begin{figure}[t]
\centering
\resizebox{0.47\textwidth}{!}{%
\includegraphics{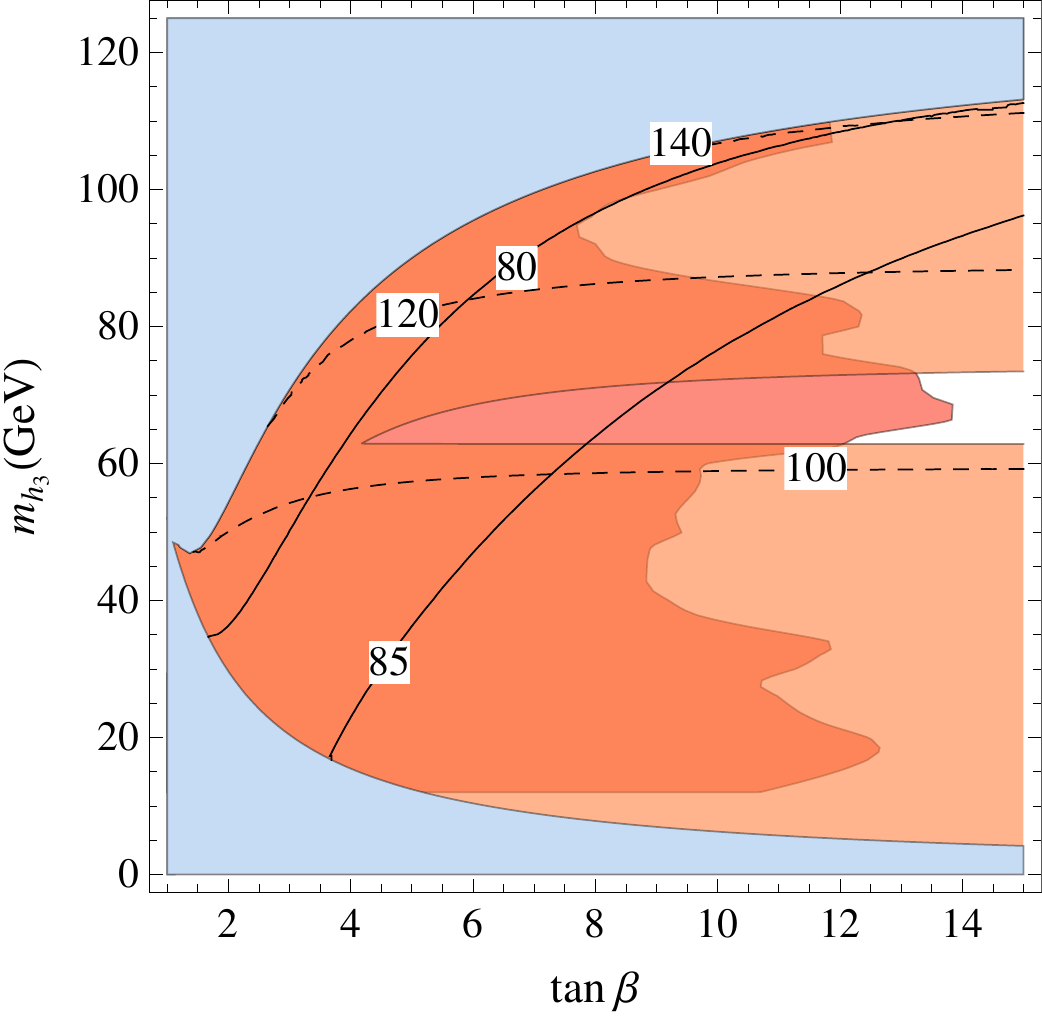}
}\hspace{.4 cm}
\resizebox{0.47\textwidth}{!}{%
\includegraphics{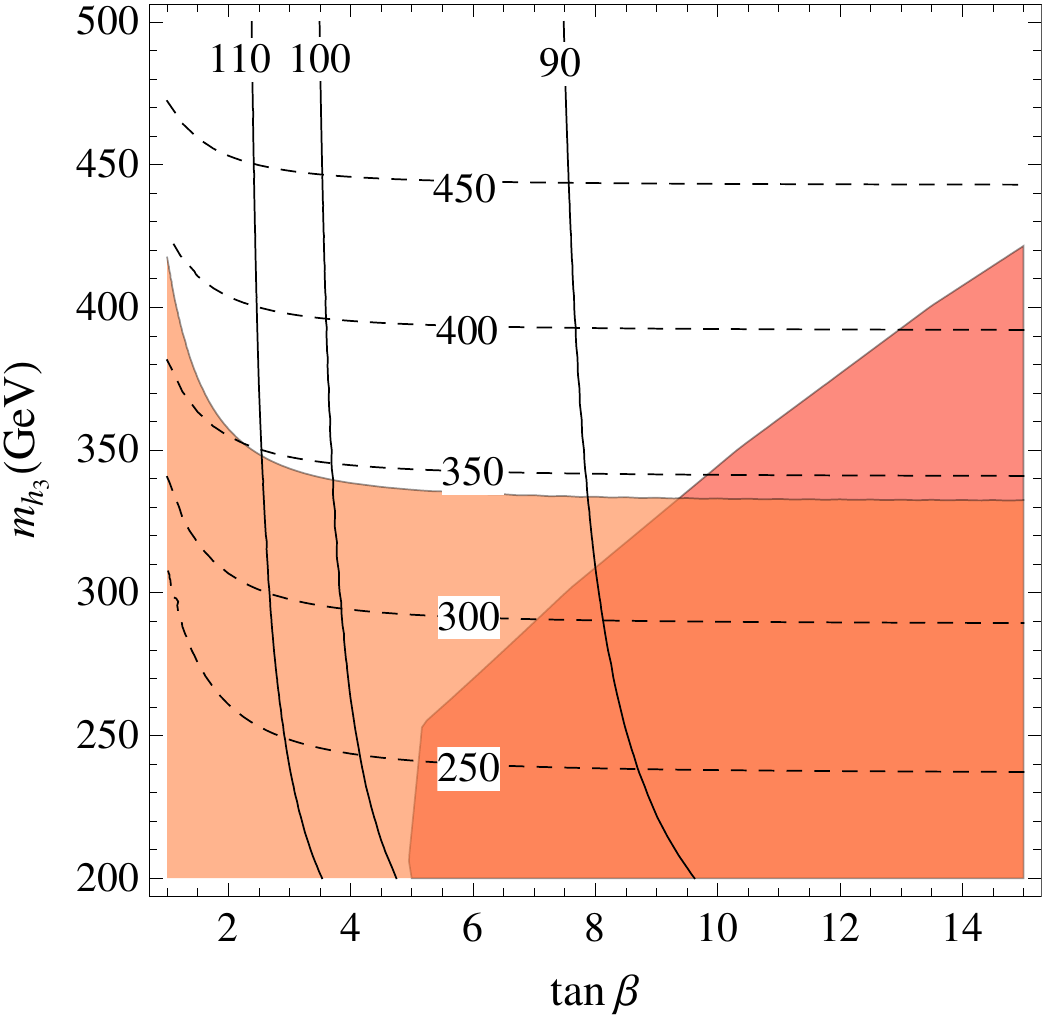}
}
\caption{Black continuous lines: values of $\Delta_t$. Red shaded areas: excluded by direct searches of $H\to b\bar{b}$ at LEP (left), and of $A,H \to \tau^+\tau^-$ at CMS \cite{CMS:gya} (right). All the rest as in fig. \ref{fig:Sdec_LHC8}. Taken from ref. \cite{Barbieri:2013nka}, where the notation $H \to h_3$ was used.}
\label{fig:MSSM}       
\end{figure}

The MSSM is a particular case of the Singlet decoupled scenario considered here, where $\lambda$ is set to zero. The measured value of $m_h$ fixes the radiative correction $\Delta_t$, so that the phenomenology of the Higgs sector depends on only two free parameters, $\tan\beta$ and $m_H$ \cite{Barbieri:2013hxa}.
In fig. \ref{fig:MSSM} we show current exclusions from direct and indirect searches, in both cases of $m_H$ larger and smaller than $m_h$. For $m_H > m_h$, the red region is excluded by CMS searches of $A,H \to \tau^+\tau^-, $\footnote{The exclusion displayed in fig. \ref{fig:MSSM} reaches a region of lower $\tan\beta$, with respect to the analogous searches displayed in fig. \ref{fig:Craig_current}. This can be explained with the fact that, to draw the exclusion in fig. \ref{fig:MSSM}, the signals in $\tau^+\tau^-$ coming from $h$, $H$ and $A$ have been added to each other, while to draw those in fig. \ref{fig:Craig_current} they have not \cite{Craig:2015jba}.
}
while for for $m_H < m_h$ it is again the LEP search for $H \to b\bar{b}$ that excludes part of the parameter space left open by the fit of the Higgs couplings. Continuous black lines indicate the value of $\Delta_t$ necessary to obtain the correct $m_h$: the higher the $\Delta_t$ the worse the fine tuning of the model.

The strategy to explore the MSSM Higgs sector is very similar to that of the NMSSM with $\phi$ decoupled, with some important differences that we summarise here
\begin{itemize}
\item[$\diamond$] the fit of the Higgs couplings does not allow for values of $m_{H,H^\pm}$ smaller than about 350 GeV. Perhaps more importantly, the LHC14 is expected to roughly double the reach in $m_{H,H^\pm}$ with Higgs coupling measurements
;
\item[$\diamond$] the region of large $\tan\beta$ and relatively small $m_H$ is now physical, and it is best probed via searches of $H,A \to \tau^+\tau^-$;\end{itemize}
%
%
We conclude with the observation that our discussion of the MSSM phenomenology holds for $\mu A_t/m_{\tilde{t}}^2 \lesssim 1$, as motivated by naturalness. By relaxing this requirement one could have, for example, alignment without decoupling (see ref. \cite{Carena:2014nza} for a recent study), but we do not explore this possibility here.

\subsection{Partial summary}
\label{sec:NMSSM_summary}
Given the current experimental information, the NMSSM appears as one of the most natural descriptions of electroweak symmetry breaking, and opens the possibility that the first particles accessible at colliders could be the scalars of the Higgs sector. With this motivation, we have outlined a possible overall strategy to probe the presence of these extra scalars. Our focus has been on the two CP-even ones $\phi$ and $H$, with a mention to direct searches for the CP-odd states.
We have first explained how to derive analytic relations among the mixing angles of the Higgses, and the six free physical parameters in (\ref{mixing_angles}) which are independent of the specific NMSSM potential. They allow for a direct and general interpretation of the impact of the Higgs couplings on the parameter space of the model.

To simplify the discussion, we have then focused on two limiting cases:
\begin{itemize}
\item[a)] doublet-like state $H$ decoupled.

The free parameters are reduced to four, which we have chosen as $m_\phi$, $\tan\beta$, the Yukawa coupling among the Higgses $\lambda$, and the radiative contribution to $m_h$, $\Delta$. Naturalness of the EW scale motivates values of $\lambda$ close to one, for this reason we have focussed on $\lambda = 1.2$ and $\lambda = 0.7$. The first value needs a UV completion before the GUT scale, while the second one stays perturbative up to that energy. The radiative correction $\Delta$ has little impact on the phenomenology, unless it becomes so small to make it difficult to obtain the correct $m_h$. Following ref. \cite{Buttazzo:2015bka}, we have then summarised prospects at current and future machines, including the future LHC stages (13 TeV, 14 TeV and HL), a 33 and a 100 TeV $pp$ colliders, as well as lepton colliders such as the ILC, CLIC, FCC-ee and CPEC.

\item[b)] singlet-like state $\phi$ decoupled.

The free parameters are reduced to three, we have chosen them as $m_H$, $\tan\beta$ and $\Delta$. In addition to the neutral CP-even state $H$, also a charged one $H^\pm$ is predicted, with a similar mass. The MSSM is just a particular realisation of this case, where the free parameters are only $m_\phi$ and $\tan\beta$; the radiative correction is in fact fixed to reproduce the measured value of $m_h$.
We have then mainly followed refs. \cite{Barbieri:2013hxa,Barbieri:2013nka,Craig:2015jba} to summarise the indirect and direct phenomenology. 
\end{itemize}
By considering these two simplified cases, one understands which signals are typical to specific CP-even scalar degree of freedom. Since such extra particles are motivated in many other extensions of the Standard Model, this analysis has a validity that, to some extent, goes beyond the NMSSM and MSSM considered here. 

The status of current searches, and the prospects for future ones, can be summarised as follows\footnote{Electroweak precision tests are not relevant to any of the above cases \cite{Barbieri:2013aza}.}
\begin{itemize}
\item[$\diamond$] \textbf{Higgs couplings measurements}


In case a) ($H$ decoupled) much of the parameter space is left unexplored by current measurements, and will be effectively probed only by a per-mille precision in the Higgs signal strengths, as the one awaited at future leptonic machines. This makes the hunt for direct signs of the new scalar $\phi$ particularly important. In case b) ($\phi$ decoupled) instead, already the LHC with 14 TeV and 300 fb$^{-1}$ is expected to probe a significant part of the model parameter space. However, a region of ``alignment without decoupling'' $\delta \simeq 0$ survives, and alternative ways to explore it are necessary.

When the lighter new scalar is the extra singlet, a sizeable deviation of the trilinear Higgs coupling from its SM value could be observable at the future LHC stages, potentially even before deviations in the Higgs signal strengths. On the contrary, if the lighter new scalar is the extra-doublet, $g_{hhh}/g_{hhh}^{\rm SM}$ is typically much closer to one.

\item[$\diamond$] \textbf{Direct searches of scalar resonances, and comparison with indirect ones}

In case a), direct searches will dominate over indirect ones in probing the parameter space of the model, in the next LHC runs. This is especially true for values of $\lambda$ below $\sim 1$, that in general look very hard to explore experimentally. More precisely, direct searches for the extra singlet, already at the LHC13, will probe a portion of parameter space at least comparable to the Higgs coupling measurements at the HL-LHC, as evident from the left-hand plots of figs. \ref{fig:NMSSM_l12} and \ref{fig:NMSSM_l07}. Future leptonic colliders will likely fill this gap, and a per-mille precision on the Higgs signal strengths will be competitive with the expected direct reach of a 100 TeV collider.
The searches of a scalar resonance relevant for the singlet-like state are those in $ZZ$ and $WW$. The $hh$ channel can play a role for certain values of the singlet vev $v_s$ and for a not too large $m_\phi$. The reason is that, for $m_\phi \gg m_W$, the equivalence theorem fixes ${\rm BR}_{\phi \to hh} = {\rm BR}_{\phi \to ZZ} = {\rm BR}_{\phi \to WW}/2$, a value for which $VV$ searches are more effective.

In case b) ($\phi$ decoupled), on the contrary, indirect searches will play a prominent role, with the exception of the region of ``alignment without decoupling''. For small values of $m_H$ and $\tan\beta$, that region is currently probed by searches of $A \to Zh$ and $H \to \tau^+ \tau^-$, as can be seen from fig. \ref{fig:Craig_current}.
In the MSSM case, direct searches are expected to dominate over indirect ones also in a region of large $\tan\beta$. Looking for scalar and pseudoscalar resonances decaying into $\tau^+ \tau^-$ and $b\bar{b}$, in fact, is already more effective than Higgs coupling measurements, as one can see in fig. \ref{fig:MSSM}.
The region of small-to-medium $\tan\beta$ is currently not probed directly, and the avenues that look more promising to do that are the $H \to t\bar{t}$ and the $H^+ \to t\bar{b}$ channels.
\end{itemize}
Higgs coupling measurements do not exclude the possibility that the extra scalar is lighter than 125 GeV. This is true for both cases of $\phi$ and $H$ decoupled, even if the possibility of $m_H < m_h$ is strongly disfavoured by ${\rm BR}_{B\to X_s \gamma}$. Direct searches are starting to explore that region, and we recommend a continuation of this effort in the next LHC runs, for example with searches of CP-even and CP-odd scalars in the $\tau^+\tau^-$ and $\gamma\gamma$ channels (see \textit{e.g.} ref. \cite{King:2014xwa} for expected signals at the LHC13).

The picture outlined above offers a useful guidance to experimental searches for signs of extra scalars. However, in case a signal --direct or indirect-- will be observed in the future, it will be necessary to interpret it in a fully-mixed situation. The relations for the mixing angles derived in \cite{Barbieri:2013hxa} provide a useful tool for this purpose, as we have shown in fig. \ref{fig:gammagamma} with an example of searches for an extra singlet $\phi$ in the $\gamma \gamma$ channel, for $m_\phi < m_h$.

\section{Flavour physics from a $U(2)^3$ symmetry}
\label{sec:flavour}


The Standard Model (SM) does not provide an understanding of the many parameters governing the flavour physics of the quark sector, namely the values of their masses and of the CKM mixing angles. Moreover, if one sticks to a natural solution of the HP, as already mentioned in the introduction one would expect to have already observed deviations from the SM predictions in the majority of CP and flavour violating observables.
To be more precise, we parametrise our ignorance as
\begin{equation}
\mathcal{L}_{\rm NP} =
\sum_i \dfrac{c_i}{\Lambda_i^2} \mathcal{O}_i,
\label{eff_L}
\end{equation}
where $\mathcal{O}_i$ are operators of dimension six that violate flavour, generated by some NP appearing at a scale $\Lambda_i$ (higher dimensional operators are suppressed by higher powers of the NP energy scale). If the coefficients $c_i$ of such operators are of order one, as expected for a generic NP flavour structure, the lower bounds on $\Lambda_i$ reach levels of $10^4$ to $10^5$ GeV (see ref. \cite{Bevan:2014cya} for an updated analysis).
This implies that, without other assumptions, one would be forced to give up on a natural solution of the hierarchy problem of the Fermi scale.
Pushing the scale of NP to those large values, included a possible solution to the SM flavour problem, would also strongly lower our hopes to test such a solution at current and future experiments.

\subsection{CKM-like symmetries and $U(2)^3$}
\label{sec:U2_model}

Insisting on the presence of some NP in the TeV range demands some mechanism to alleviate the strong flavour bounds that we just mentioned. Motivated by the fact that measurements are consistent with a CKM picture to a level of $20\div 30 \%$, we consider the specific mechanisms that give rise to a CKM-like structure of flavour and CP violation, defined as follows.
Only the flavour violating effective operators $\mathcal{O}$ that are present also in the SM are allowed, with the respective Wilson coefficients proportional in size to the SM ones, \textit{i.e.} to products elements of the CKM matrix $V$.
This means that the coefficients $c_i$ of eq. (\ref{eff_L}) have the form
\begin{equation}
c_i \sim C_i \,(V_{3a} V_{3b})^{1\div2}\,,
\label{CKM_coefficients}
\end{equation}
with $C_i$ of order one, and with details of the CKM elements depending on the operator under consideration.
For example, a term in $\mathcal{L}_{\rm NP}$ relevant for $K-\bar{K}$ mixing has the form $C_{LL}^K (V_{ts} V_{td}^*)^2\,(\bar{d}_L\gamma_\mu s_L)^2/2$.
The operators that are absent in the SM, like for example $(\bar{s}_R d_L)\,(\bar{s}_L d_R)$, have instead to be very suppressed. We refer the reader to ref. \cite{Barbieri:2014tja} for a complete list of the allowed operators, and their size.
For generality and simplicity, we stick to those models that realise a CKM-like picture of flavour thanks to (continuous) flavour symmetries only.

It appears \cite{Barbieri:2014tja} that the only two symmetries that realise a CKM-like pattern, as just defined, are $U(3)^3 = U(3)_q \times U(3)_u \times U(3)_d$ \cite{Chivukula:1987py,Hall:1990ac,D'Ambrosio:2002ex} (or equivalent), broken by \footnote{A flavour symmetric model is defined not only by the choice of the symmetry, but also by its breaking directions (``spurions''). In the NP Lagrangian all the operators, that one can build using the field content of the model plus the spurions, are demanded to respect such a symmetry, so that an example of operator within $U(3)^3$ is $(\bar{q}_L \,\gamma_\mu\,Y_u Y_u^\dagger\,q_L)^2$. The rotation to the mass basis then gives rise to coefficients proportional to products of CKM matrix elements.} 
\begin{equation}
Y_u = (3, \bar{3}, 1)\,, \; Y_d = (3, 1, \bar{3}),
\label{eq:U3spurions}
\end{equation}
and  $U(2)^3 = U(2)_q \times U(2)_u \times U(2)_d$ \cite{Barbieri:2011ci,Barbieri:2012uh} (or equivalent, see also \cite{Feldmann:2008ja,Kagan:2009bn}), broken by
\begin{equation}
\Delta_u = (2, 2, 1)\,, \; \Delta_d = (2, 1, 2)\,, \; V = (2, 1, 1)\,.
\label{eq:U2spurions}
\end{equation}
The framework defined by $U(3)^3$ and eq. (\ref{eq:U3spurions}) is often named in the literature as ``Minimal Flavour Violation'' (MFV).
Next to the spurions in eqs. (\ref{eq:U3spurions}) and (\ref{eq:U2spurions}) we have indicated, in parenthesis, their transformation properties under the respective group. We have not been able to construct a symmetry, non-equivalent to one of the two above, giving a CKM-like pattern. One of the reasons is that, in order to maintain this pattern, one has to introduce some distinction between the left and right quarks, as well as between $u_R$ and $d_R$ (to avoid chirality breaking operator). See section 3 of ref. \cite{Barbieri:2014tja} for a more thorough discussion of this point, together with explicit examples of symmetries that do not give a CKM-like picture.

\subsection{Status of CKM-like symmetries in meson mixing}
\label{sec:U2_pheno}
\begin{figure}[t]
\centering
\resizebox{0.8\textwidth}{!}{%
\includegraphics{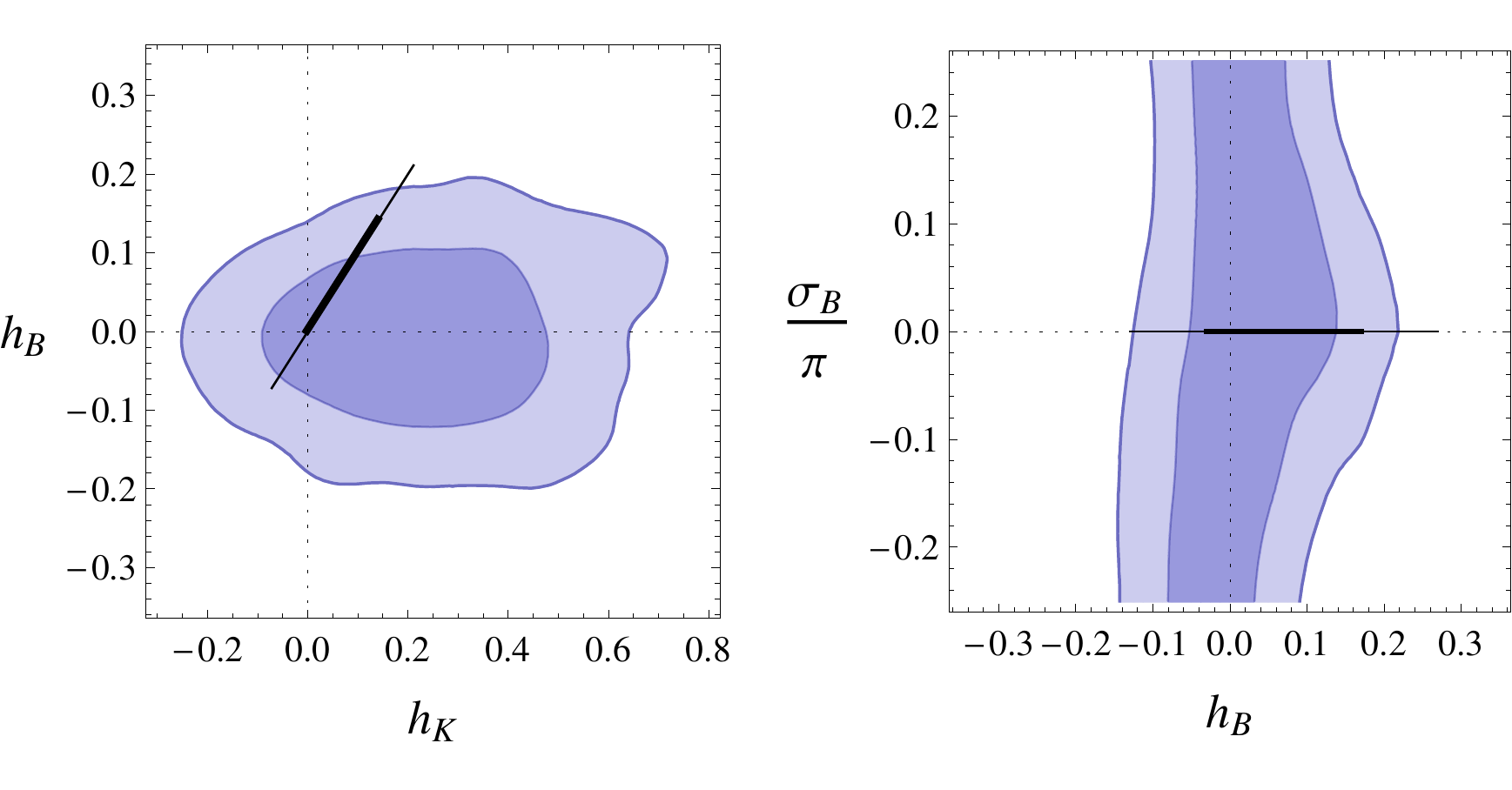}
}

\caption{Results of the $U(2)^3$ CKM fit. The dark and light -shaded regions correspond to 68\% and 95\% probability, respectively. The segments show the corresponding results for MFV, the thick and thin lines corresponding to 68\% and 95\% probability. Taken from ref. \cite{Barbieri:2014tja}.}
\label{fig:DF2plots1}
\end{figure}

The mixing amplitudes relevant for the $K$, $B_d$ and $B_s$ mesons can be written, in a fully general way, as
\begin{equation}
M_{12}^K = (M_{12}^K)_{\rm SM} \left( 1+ h_K e^{2i\sigma_K}\right) , \qquad 
M_{12}^{d,s} = (M_{12}^{d,s})_{\rm SM} \left( 1+ h_{d,s} e^{2i\sigma_{d,s}}\right).
\label{eq:Mixing_Amplitudes}
\end{equation}
The two frameworks predict
\begin{align}
&\sigma_K = 0, \qquad \sigma_d = \sigma_s \equiv \sigma_B, \quad  &\quad h_d = h_s \equiv h_B\,, & &\qquad U(2)^3 \\
&\sigma_K = \sigma_d = \sigma_s = 0 \quad  &\quad h_K = h_d = h_s \equiv h\,, & &\qquad U(3)^3
\end{align}
so that the phenomenology of $U(3)^3$ has less space for deviations from the SM than the one of $U(2)^3$. In particular, $U(3)^3$ predicts the ratios of analogous $K$, $B_d$ and $B_s$ observables to be SM-like, and all phases to be aligned with the CKM ones. In $U(2)^3$ just $B_d$ and $B_s$ observables are correlated, and only $K$ ones are predicted to be aligned in phase with the SM.
We do not mention here up-quark nor observables that violate flavour number $F$ by one unit ($\Delta F = 1$) since, with the possible exception of $\epsilon^\prime/\epsilon$ \cite{Barbieri:2012bh}, they are presently less constraining than the $\Delta F = 2$ ones (like $\epsilon_K$, $\Delta M_{d,s}$, $\phi_{d,s}$) that we consider \cite{Barbieri:2012uh}.

The results of the global fit of ref. \cite{Barbieri:2014tja}, where we varied the CKM Wolfenstein parameters together with $h \; (U(3)^3)$ and $h_K, h_B, \sigma_B \; (U(2)^3)$ are shown in Fig. \ref{fig:DF2plots1}. Flavour measurements are then probing scales of $4\div7$ TeV. To understand this, it is sufficient to know that the phenomenological $h_i$ parameters satisfy the relation
\begin{equation}
h_i \simeq C_i \left(\frac{3.1\, {\rm TeV}}{\Lambda_i} \right)^2\!,
\end{equation}
where we remind that the $C$'s are the Wilson coefficients of the relative effective operators, expected to be of order one since the CKM suppressions is factored out. Fig. \ref{fig:DF2plots1} also quantifies our previous statement that $U(2)^3$ (blue regions) allows for larger deviations from the SM than $U(3)^3$ (black lines).
In case a NP signal will show up in some flavour observable, it could then be possible to distinguish the two symmetries. For example the observation of a $\sigma_B \neq 0$ would strongly disfavour $U(3)^3$, but would be welcome in $U(2)^3$. Deviations in $B_d$ and $B_s$ observables, if not correlated, would be in tension with both symmetries.


\subsection{Flavour vs collider searches in Supersymmetry}
\label{sec:U2_SUSY}

When $U(3)^3$ and $U(2)^3$ are implemented in Supersymmetry, do collider bounds on sparticle masses allow to expect deviations from the SM, at near future flavour experiments? 

To answer this question, in ref. \cite{Barbieri:2014tja} we have considered a supersymmetric spectrum whose main feature is that squarks of the first two generations are heavier than all the other sparticles, as motivated by naturalness. Notice that such a spectrum is easily defendable within a $U(2)^3$ framework, since $U(2)^3$ distinguishes the first two generations from the third one. We have then performed a numerical analysis of some relevant flavour violating processes, including carefully all the loop contributions (Gluinos, Winos, Binos, Higgsinos and Charged Higgs), which we also studied analytically to have a control on the scan. To perform it and compute flavour observables, we used SUSY\_FLAVOR 2.10 \cite{Crivellin:2012jv} as well an independent program developed by us, finding very good agreement. We imposed all the bounds from direct searches available at the end of the LHC8 run, and we refer the reader to ref. \cite{Barbieri:2014tja} for further details on the scan.
\begin{figure}[t]
\centering
\resizebox{0.47\textwidth}{!}{%
\includegraphics{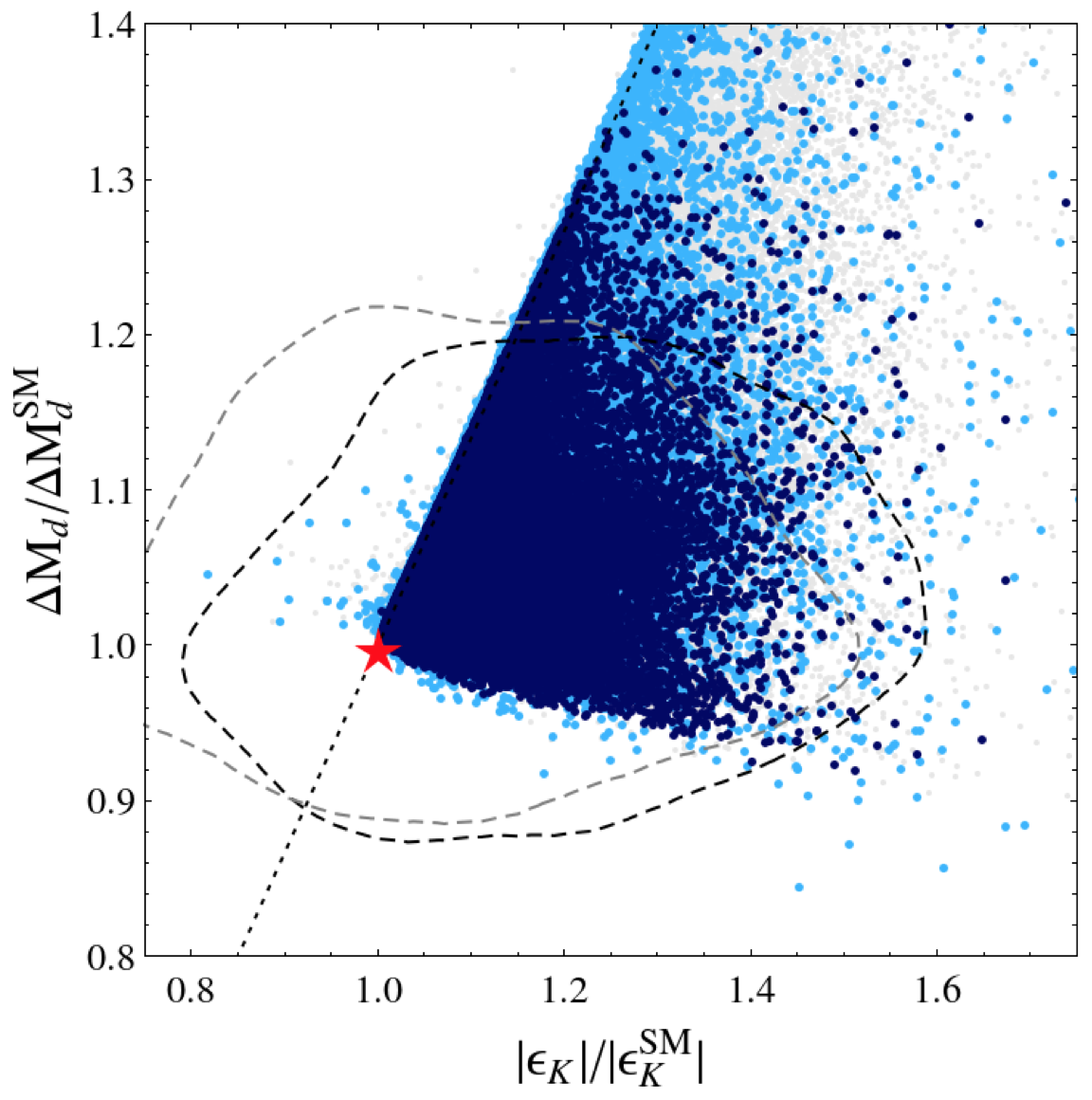}}\hspace{.4 cm}
\resizebox{0.49\textwidth}{!}{%
\includegraphics{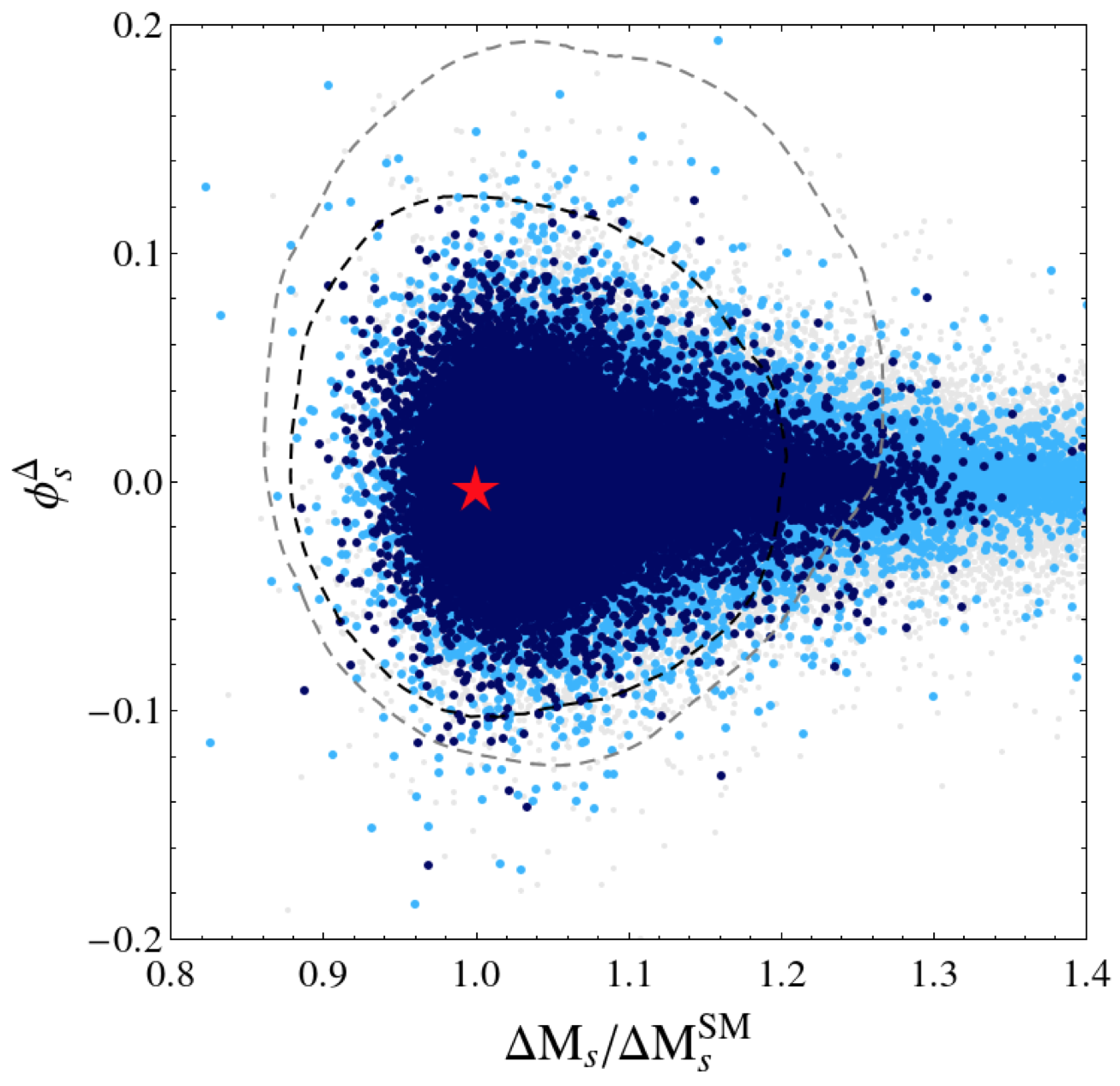}}
\caption{Correlation between the $\Delta F=2$ observables $|\epsilon_K|$ and $\Delta M_d$ (left), $\Delta M_s$ and $\phi_s^\Delta$ (right). Light and dark blue points are allowed by all constraints, light blue points have a compressed spectrum (see Section 5.2 of Ref. \cite{Barbieri:2014tja} for more details), light gray points are ruled out by $B \to X_s \gamma$. The red star is the SM. The black dashed lines are the 95\% C.L.\ regions allowed by the global CKM fit imposing the $U(2)^3$ relations, the gray dashed lines are the same without imposing the $U(2)^3$ relations. The dashed line in the left hand plot shows the MFV limit. Taken from ref. \cite{Barbieri:2014tja}.}
\label{fig:scatter} 
\end{figure}

We show in Fig. \ref{fig:scatter} the predictions that we obtain as described above, within the $U(2)^3$ framework. The MFV limit reduces the figures to the dashed line in the left-hand plot, and to an horizontal $\phi_s = 0$ line in the right-hand plot. Among the $\Delta F = 2$ observables that we show, the most promising ones for the detection of possible deviations are (we base our conclusion on the future expectations given in ref.~\cite{Charles:2013aka}):
\begin{itemize}
\item[$\circ$] $\phi_s$, because LHCb is awaited to reach a precision of a few per-mille in its measurement, which is also the level of the theoretical uncertainty in the SM prediction \cite{Charles:2011va};
\item[$\circ$] $\Delta M_{d,s}$, mainly because of expected improvement in lattice QCD, which could push the current 20-30\% constraint to the level of 5\%.
\end{itemize}
Concerning the limits on CKM-like scenarios from $\epsilon_K$, the prospects for improvement depend mostly on future lattice computations of long-distance contributions~\cite{Ligeti:2016qpi}. Concerning $D-\bar{D}$ mixing, we do not expect deviations from the SM at an observable level at near-future experiments, as in the case of $U(2)^3$ without a SUSY embedding \cite{Barbieri:2012uh}.

\begin{figure}
\centering
\resizebox{0.485\textwidth}{!}{%
\includegraphics{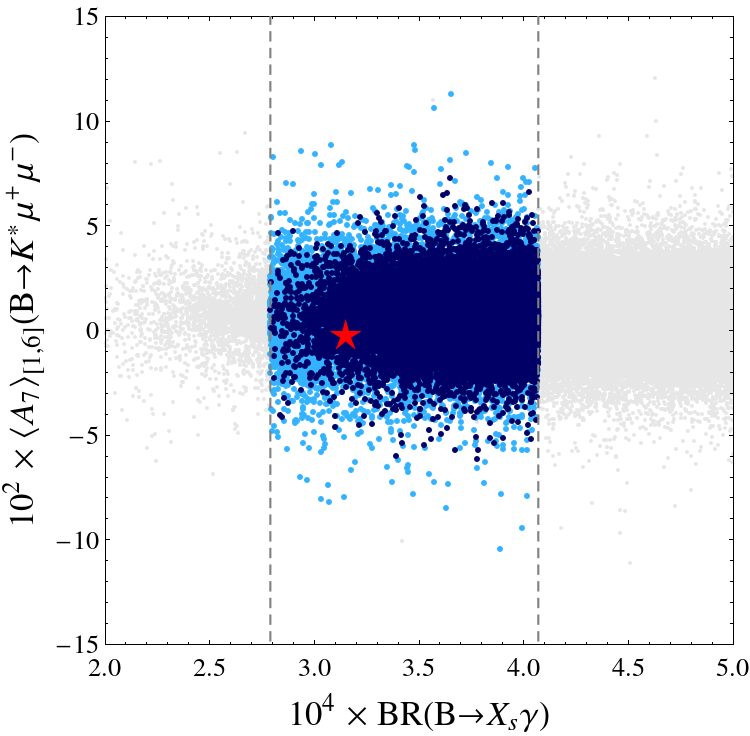}}\hspace{.4 cm}
\resizebox{0.48\textwidth}{!}{%
\includegraphics{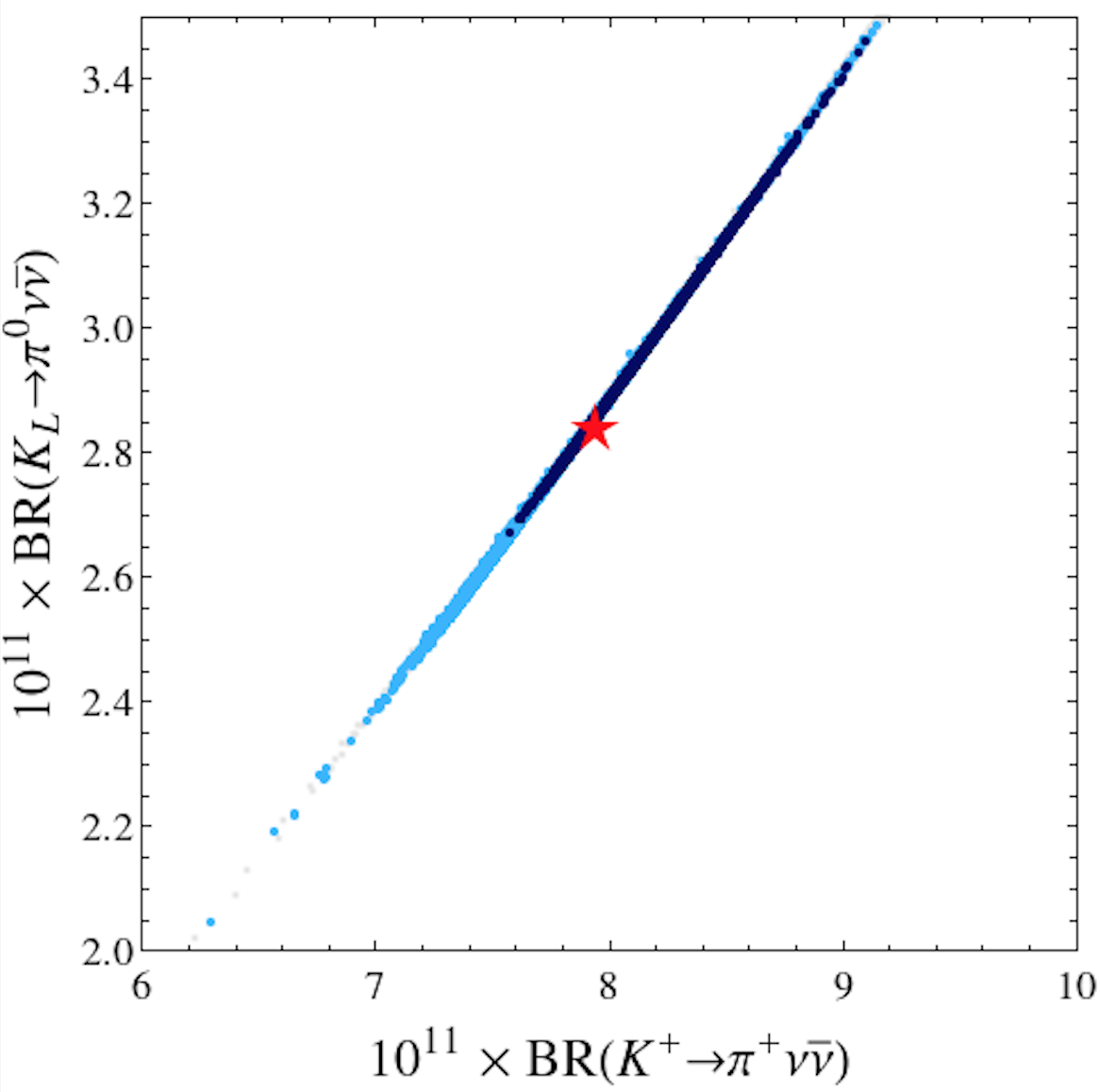}}

\caption{Correlation between the $\Delta F=1$ observables BR$(B\to X_s\gamma)$ and $\langle A_7\rangle_{[1,6]} (B \to K^* \mu^+\mu^-)$ (left), and BR$(K^+\to \pi^+ \nu \bar{\nu})$ and BR$(K_L\to \pi^0 \nu \bar{\nu})$ (right). Points and star as in fig. \ref{fig:scatter}. Left: taken from \cite{Barbieri:2014tja}. Right: courtesy of D.M. Straub (see also footnote~\ref{foot:Kpinunu}).}
\label{fig:scatter_C7}
\end{figure}

Concerning $\Delta F = 1$ processes, the dominant constraint is the one from the branching ratio of $B \to X_s \gamma$ (that is respected by all light and dark -blue points).
Some $\Delta F = 1$ observables are very sensitive to the value chosen for $\tan\beta$, and the fact that we limited the scan to $\tan \beta < 5$ has relevant phenomenological consequences, for example it keeps deviations from the SM in ${\rm BR}_{B_{d,s} \to \mu^+ \mu^-}$ below the 30\% level. Measurements where one can hope to see NP signs from $U(2)^3$ are angular CP asymmetries in $B \to K^* \ell^+ \ell^-$, like $A_7$, and the branching ratio of $K \to \pi \nu \bar{\nu}$:\footnote{\label{foot:Kpinunu}
Notice that the SM central values for the right-hand plot of fig. \ref{fig:scatter_C7} have shifted to BR$(K^+ \to \pi^+\nu\bar{\nu}) = 9.11 \pm 0.72 \times 10^{-11}$ and BR$(K^+ \to \pi^+\nu\bar{\nu}) = 3.0 \pm 0.3 \times 10^{-11}$ \cite{Buras:2015qea}. What matters for our discussion is however the relative spread due to NP, which does not change.}
both are shown in fig. \ref{fig:scatter_C7}. Analyses of the former observables are awaited from LHCb, for which a possible precision goal can be that of a few per-cent, necessary to probe the supersymmetric realisation of $U(2)^3$. The NA62 experiment at CERN has just started its pilot run, and it aims to reach a sensitivity of $10 \%$ in ${\rm BR}_{K^+ \to \pi^+\nu\bar{\nu}}$ \cite{Romano:2014xda}, thus providing interesting complementary information with respect to direct LHC searches.
Other deviations from the SM could be observed in the mixing-induced CP asymmetries in the decays $B \to \phi K_S$ and $B \to \eta' K_S$, correlated in a specific way with the effects described above \cite{Barbieri:2011fc}.

\subsection{Partial summary}
\label{sec:U2_summary}
Which impact will flavour measurements have on the medium-term future of particle physics?
We have tried to address this question in the spirit of ref. \cite{Barbieri:2014tja},  assuming New Physics to be present in the TeV range and considering flavour symmetric models as a general solution to the NP flavour problem. We have focused on $U(3)^3$ and $U(2)^3$, as the only two giving raise to a CKM-like structure of flavour and CP violation, which is respected to a very good level in Nature. Within these models, current flavour measurements are probing scales up to $4 \div 7$ TeV, and $U(2)^3$ allows for a richer pattern of deviations from the SM than $U(3)^3$. We also studied the interplay of flavour measurements with collider searches in the specific case of Supersymmetry, with a spectrum motivated by naturalness.

Sizeable deviations from the SM in flavour observables, especially in $\phi_s$ (only in $U(2)^3$) and $\Delta M_{d,s}$ and $\epsilon_K$, are possible, and we are eagerly waiting for related progresses in the experimental and lattice QCD fields.
Deviations from the SM are also expected in $\Delta F = 1$ observables, as ${\rm BR}_{K^+ \to \pi^+\nu\bar{\nu}}$ and CP asymmetries in semileptonic ($A_7$) and fully hadronic ($S_{\phi K_S}$, $S_{\eta' K_S}$) $B$ decays. Such deviations will likely be accompanied by sparticles within future LHC reach, and the pattern of correlations predicted by $U(2)^3$ will help in distinguishing it from different flavour models.

We finally comment on the recent anomalies observed in semileptonic $B$ decays, pretending they are not due to underestimated SM effects (see ref. \cite{Altmannshofer:2015sma
} for an updated analysis). Both the kind and the size of the NP needed to solve the tensions is, in general, in the ballpark of what expected in a $U(2)^3$-symmetric framework. However, the possibility to address the anomalies is challenged when $U(2)^3$ is embedded in Supersymmetry, because of the extra suppressions (coupling constants, loop functions) typical of this picture. Concerning deviations from lepton universality in $B$ decays, no definite conclusion can be drawn in $U(2)^3$, unless some modelling for the lepton sector is specified.

\section{Conclusions}
\label{sec:conclusions}
The lack of evidence for New Physics at the energy and intensity frontiers is challenging our beliefs about a deeper theory of Nature. In particular, it induces us to rethink  the naturalness of the Fermi scale, which is the only general motivation for deviations from the Standard Model at accessible energy scales. This rethinking, while certainly a much needed exercise, has not (yet) provided indications for an energy scale at which to aim, at least not in a very general way. The standard approach to the hierarchy problem, then, still provides the main tool to orientate future experimental efforts of the community.

With this in mind, we have focused on the Higgs and flavour sectors of the Standard Model. They are possibly the two areas where natural NP is expected to show up first. What is the impact of Higgs and flavour measurements, as a whole, on the future of particle physics? How do these precision measurements interplay with direct searches for new particles?
We have reviewed the answers to these questions in a few of the most natural NP models, giving a particular attention to recent and future experimental developments.
Concerning the Higgs, a quite general property of natural models is the presence of extra ``Higgs bosons''. Supersymmetry is the prototype of theories providing new uncoloured scalars, and in the NMSSM the possibility that they are the lightest new particles is also favoured --to some extent-- by naturalness arguments. We have discussed the Higgs sector of the NMSSM, as well as that of the MSSM, in sect. \ref{sec:NMSSM}.
Coming to flavour, data constrain deviations from the CKM picture to lie below the 20-30\% level, thus motivating frameworks that realise such a picture. Based on flavour symmetries only, one can obtain a CKM-like description of flavour with either a $U(3)^3$ or a $U(2)^3$ symmetry of the quark sector. We have discussed them in sect. \ref{sec:flavour}, both in the general case and in the specific scenario of Supersymmetry. We have given more attention to $U(2)^3$, both because it allows for larger deviations from the SM in the near future, and because it accommodates more easily NP that combines naturalness with collider constraints (\textit{i.e.} a NP scale associated with the first two generations heavier than that associated with the third one).

We refer to the partial summaries of the Higgs and flavour parts, sects. \ref{sec:NMSSM_summary} and \ref{sec:U2_summary}, for a schematic outline of the respective phenomenologies. The reader can find there the observables for which larger deviations from the SM are expected. We hope that the synthetic fashion of the exposition can render this paper a useful phenomenological orientation, especially for experimentalists.


\subsection*{Acknowledgements}
I warmly thank
Riccardo Barbieri, Dario Buttazzo, Andrea Tesi, David M. Straub, and Kristjan Kannike for the stimulating and pleasant collaborations on the subjects presented in this paper. I thank Andrea Tesi for a careful read of the manuscript, and David Straub for useful discussions about the flavour part
. I am also grateful to the INFN for the Fubini prize awarded for my Thesis \cite{Sala:2013hga}, which lead to this publication.
This work is supported by the European Research Council ({\sc Erc}) under the EU Seventh Framework Programme (FP7 2007-2013)/{\sc Erc} Starting Grant (agreement n.\ 278234 --- `{\sc NewDark}' project). The final publication is available at Springer via \href{http://dx.doi.org/10.1140/epjp/i2016-16079-5}{http://dx.doi.org/10.1140/epjp/i2016-16079-5}.

\end{document}